\documentclass[preprint,12pt,authoryear]{elsarticle}
\usepackage{amssymb}
\usepackage{amsmath}

\usepackage{cite}

\usepackage{algorithm}
\usepackage{algorithmicx}
\usepackage{algpseudocode}
\usepackage[ruled, vlined, algo2e]{algorithm2e}

\usepackage{hyperref}
\hypersetup{
hypertex=true,
colorlinks=true,
linkcolor=blue,
anchorcolor=blue,
citecolor=blue
}
\usepackage{multirow}
\usepackage{multicol}
\usepackage{makecell}
\usepackage{booktabs}

\usepackage{lineno}

\usepackage{xurl}

\journal{Expert Systems with Applications}

\begin{document}
\begin{frontmatter}

\title{Encoder-Inverter Framework for Seismic Acoustic Impedance Inversion} 
\author[label1,label2]{Junheng Peng}
\author[label1,label2]{Yingtian Liu} 
\author[label2]{Xiaowen Wang}
\author[label1,label2]{Yong Li}
\author[label1,label2]{Mingwei Wang}

\affiliation[label1]{organization={The Key Laboratory of Earth Exploration \& Information Techniques of Ministry Education},
            addressline={Chengdu University of Technology}, 
            city={Chengdu},
            postcode={610059}, 
            state={Sichuan},
            country={China}}
\affiliation[label2]{organization={School of Geophysics},
            addressline={Chengdu University of Technology}, 
            city={Chengdu},
            postcode={610059}, 
            state={Sichuan},
            country={China}}

\begin{abstract}
Seismic acoustic impedance inversion is a challenging problem in geophysical exploration, primarily due to the scarcity of well-logging data and the inherent nonlinearity of the task. Most existing inversion methods, including semi-supervised learning approaches, still face limitations in accuracy and robustness. In this work, we propose a novel Encoder-Inverter framework that maps continuous seismic traces into high-dimensional linear features, thereby transforming the inversion task into a linear extrapolation or interpolation problem to enhance stability and performance. To achieve this, we introduce two auxiliary models to assist in encoder training and adopt a heterogeneous model structure to prevent shortcut learning, enabling the extraction of more generalizable and effective linear features. We evaluate the proposed method on widely used benchmark datasets, and experimental results demonstrate that our approach achieves superior inversion accuracy and robustness compared to previous methods. To promote reproducibility, we will also open-source the data and code.
\end{abstract}

\begin{highlights}
\item Acoustic impedance is an important parameter for characterizing subsurface media.
\item The high cost and scarcity of wells make it difficult to obtain the acoustic impedance of the entire profile.
\item Propose feature extraction and fine-tuning famework for seismic acoustic impedance inversion. 
\item Mapping seismic traces into linear features can effectively addresses the challenge of sparse well-logging data.
\item The proposed framework achieves good performance, robustness, and efficiency.
\end{highlights}


\end{frontmatter}


\section{Introduction}

Seismic acoustic impedance inversion (AII) is a fundamental yet challenging task in geophysical exploration, enabling the imaging of subsurface lithology and structural variations using seismic data and limited well-logging information \citep{1, 2}. However, due to the high cost of exploration, well-logging data are often scarce, rendering the AII problem ill-posed \citep{3}.

Early research on AII primarily focused on optimization-based methods, such as least squares \citep{4}, simulated annealing \citep{5}, ant colony algorithms \citep{6}, genetic algorithms \citep{7}, and particle swarm optimization \citep{8}. Probabilistic modeling approaches, including iterative Bayesian methods \citep{60, 61}, have also been explored. Although these methods have become well established over the past decades \citep{9, 10}, several challenges remain: (1) their performance is highly dependent on the initial model \citep{55}, which is often difficult to obtain; (2) they require accurate seismic wavelets, making wavelet extraction a prerequisite; (3) they generally suffer from low computational efficiency; and (4) the lack of prior information and reliance on manually designed regularization terms (such as $L_1$ and $L_p$ norms) limit their ability to effectively characterize subsurface media.

\begin{figure}
\centering
\noindent\includegraphics[width=\textwidth]{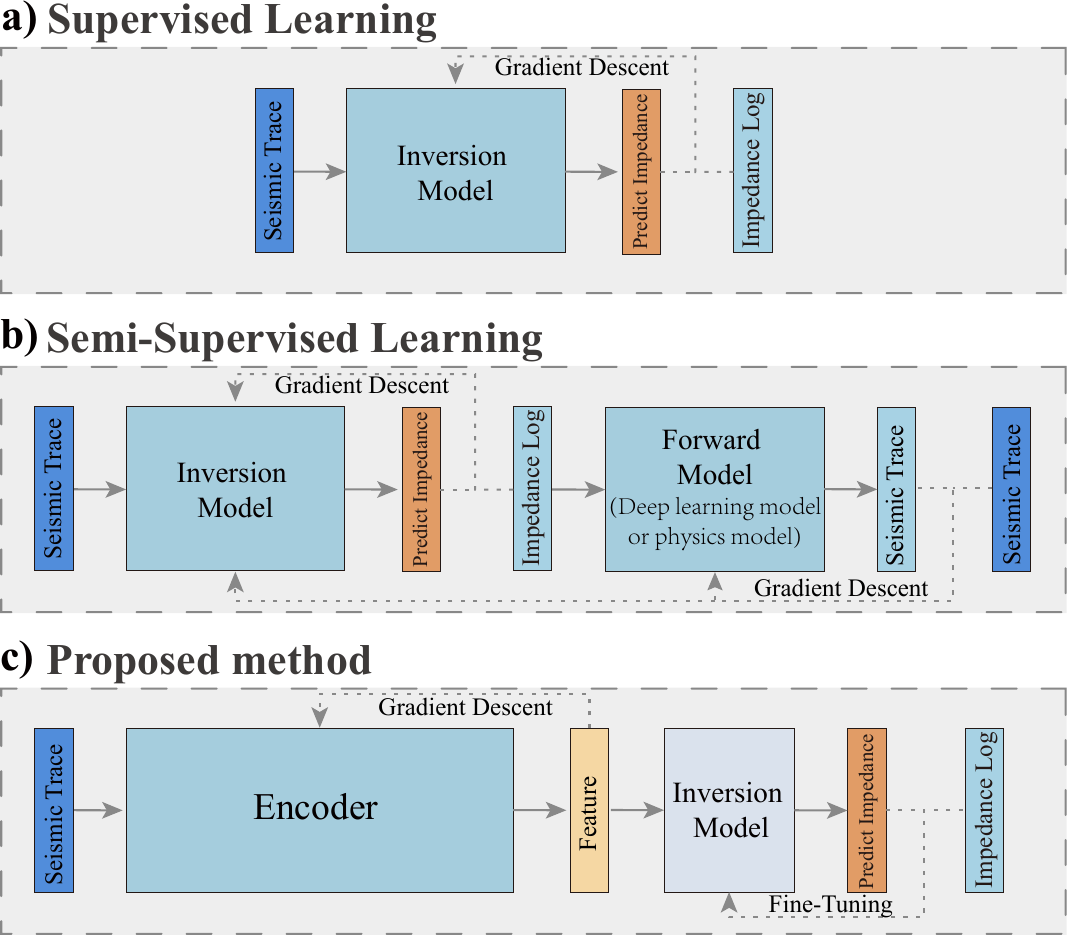}
\caption{a) supervised learning framework; b) semi-supervised learning framework; c) proposed method based on feature extraction and fine-tuning.}
\label{introduction}
\end{figure}
With the advancement of computer technology, deep learning (DL) methods have provided new solutions for seismic exploration tasks, including seismic noise attenuation \citep{11, 12}, fault identification \citep{28, 71}, seismic interpolation \citep{73}, seismic facies recognition \citep{15, 72}, seismic wavefield evolution \citep{27}, and seismic inversion \citep{16, 17, 74}. \citet{18} first introduced convolutional neural networks (CNNs) to AII, laying the foundation for subsequent DL developments in this field. DL approaches can effectively address ill-posed nonlinear inversion problems and do not require prior knowledge of physical models \citep{64}. Recent research on AII has primarily focused on DL, with significant structural innovations that can be broadly categorized into two stages, as illustrated in Figure~\ref{introduction}a and b. Early DL models for AII were based on supervised learning strategies \citep{18, 19, 63}. These models establish a nonlinear mapping between seismic traces and acoustic impedance using labeled data (acoustic impedance well-logging data), and directly apply this mapping to unlabeled seismic traces for AII \citep{25}. \citet{21} proposed the use of residual networks (ResNet) for supervised AII; \citet{22} introduced multi-scale supervised AII to ensure energy consistency of seismic signals across different frequency bands; \citet{23} incorporated the self-attention mechanism into AII. However, supervised learning often fails to achieve optimal performance, primarily due to the scarcity of well-logging data, which is limited by high acquisition costs.

AII can be viewed as a few-shot DL problem \citep{24} due to the limited availability of impedance well-logging data for supervised learning (traces with well-logging data typically account for only about 1$\%$ to 5$\%$ of the total seismic traces). To address this challenge, many researchers have proposed semi-supervised learning methods \citep{31, 56, 62}, as illustrated in Figure~\ref{introduction}b. The core idea of semi-supervised learning is to incorporate a forward model into the original supervised framework. For the inversion results of unlabeled traces, the forward model provides additional constraints on the inversion model \citep{26}. Semi-supervised DL methods can be broadly categorized into two types: those that use a physics-driven forward model \citep{32}, and those that employ a data-driven forward model \citep{33}. Physics-driven forward models face challenges similar to traditional approaches, such as the requirement to extract seismic wavelets during forward modeling. These models typically rely on seismic wavelets, followed by the forward synthesis of seismic data \citep{34}. To overcome the limitations associated with wavelet extraction, many authors have proposed using data-driven DL models as alternatives \citep{35, 37, 56}. At this stage, DL methods for AII have established a relatively mature framework, within which various improvements have been made. For example, \citet{38} proposed an attention mechanism-based dual-branch inversion model capable of extracting multi-scale sequential features; \citet{39} introduced a fusion neural network that can simultaneously invert both porosity and acoustic impedance to enhance inversion performance; \citet{40} developed a semi-supervised DL algorithm with iterative gradient correction to improve model robustness. Additionally, it is worth noting that most DL methods still employ initial models to enhance inversion effectiveness \citep{40, 41}.

Due to the extremely small proportion of labeled traces (seismic traces containing well-logging data), AII consistently faces the challenges of extrapolation and interpolation, making it critical to enhance the extrapolation and interpolation capabilities of DL models. To address this, we propose a DL method based on feature extraction and fine-tuning, as illustrated in Figure~\ref{introduction}c. The core idea is to train a high-performance encoder that effectively maps seismic traces into high-dimensional linear features. By linearizing the relationship between labeled and unlabeled seismic traces, the complexity of processing unlabeled traces is significantly reduced. Similar concepts have been successfully validated in previous studies \citep{47, 48}. To realize this, we introduce a Dimension Reducer (with fewer than 1k learnable parameters) to assist in encoder training. Additionally, we incorporate a Reconstructor to prevent the encoder from diverging by reconstructing the extracted features back into seismic data. Finally, the well-trained encoder and a small amount of impedance well-logging data are used to fine-tune the inversion model. To avoid shortcut learning during encoder training \citep{51}, both the Reconstructor and the inversion model adopt architectures different from that of the encoder. Since the proposed method does not involve a forward model, it does not require seismic wavelet extraction or forward model training, and can achieve good results without an initial model. We conducted experiments on multiple datasets using consistent hyperparameters, and the results demonstrate that the proposed method achieves superior robustness and performance compared to supervised and semi-supervised methods. Furthermore, we have open-sourced the code and data to support future DL research in the AII field.

\section{Methods}
\subsection{Forward modeling}
Seismic acoustic impedance is a key parameter for characterizing subsurface reservoirs and is widely used in oil and gas exploration. It is defined as
\begin{equation}
I = V_p \times D,
\end{equation}
where $V_p$ denotes the P-wave velocity and $D$ denotes the density. Acoustic impedance ($I$) can only be directly measured through well logging, which is expensive and results in limited data availability.

Seismic records provide indirect information about acoustic impedance. Typically, a seismic trace is modeled as the convolution of the reflection coefficient series with a seismic wavelet \citep{42}:
\begin{equation}
S(t) = R(t) \ast W(t) + n(t),
\label{Seismic Synthesis}
\end{equation}
where $t$ is the sampling time, $S$ is the seismic trace, $R$ is the reflection coefficient series, $W$ is the seismic wavelet, and $n$ represents additive random noise, which can generally be mitigated using noise attenuation techniques. The reflection coefficient series can be calculated from acoustic impedance as follows:
\begin{equation}
R(t) = \frac{I_{t+\Delta t} - I_t}{I_{t+\Delta t} + I_t}.
\label{Reflection Coefficient}
\end{equation}
By combining Equations~\ref{Seismic Synthesis} and~\ref{Reflection Coefficient}, seismic traces can be synthesized from acoustic impedance. However, while the forward modeling process is relatively straightforward, the inverse problem of recovering acoustic impedance from seismic data is highly challenging.

\subsection{Encoder-Inverter framework}

AII is inherently ill-posed, and traditional optimization methods are constrained by their dependence on accurate initial models and seismic wavelets. This limitation has led to the widespread adoption of deep learning (DL) approaches. However, most existing works primarily focus on enhancing model generalization, often overlooking the intrinsic characteristics of the seismic data itself. Consider $x$ consecutive seismic traces $\{ S_1(t), S_2(t),...S_x(t)\}$, whose underlying features are governed by a nonlinear function $S_x(t) = f(x,t)$ that is difficult to explicitly characterize. The nonlinearity of $f$ complicates the relationships among $S_x$, making interpolation and extrapolation in AII particularly challenging. To address this, we propose two key innovations: first, we decompose $f$ into a high-dimensional linear function using an Encoder; second, we construct the Inverter with a model based on linear layers, as multilayer perceptrons exhibit strong linear extrapolation capabilities \citep{49}.

Generally, linear problems are easier to solve than nonlinear ones \citep{45}. For the Inverter, when the input seismic trace features are linearized and supervised learning is performed with a small number of labeled traces, the AII problem for unlabeled traces is transformed into a linear extrapolation or interpolation task, thereby reducing its complexity. To achieve this, we introduce two auxiliary models (the Dimension Reducer and Reconstructor) to assist in training the Encoder, as detailed in the following sections. After the Encoder is trained, a small amount of well-logging data is used to fine-tune the Inverter. Throughout the entire process, well-logging data is only utilized during the fine-tuning of the Inverter. Furthermore, the proposed Inverter adopts a streamlined structure, enabling excellent inversion results without the need for additional forward models or initial models.

\subsection{Structure of models}
\begin{figure}
\centering
\noindent\includegraphics[width=\textwidth]{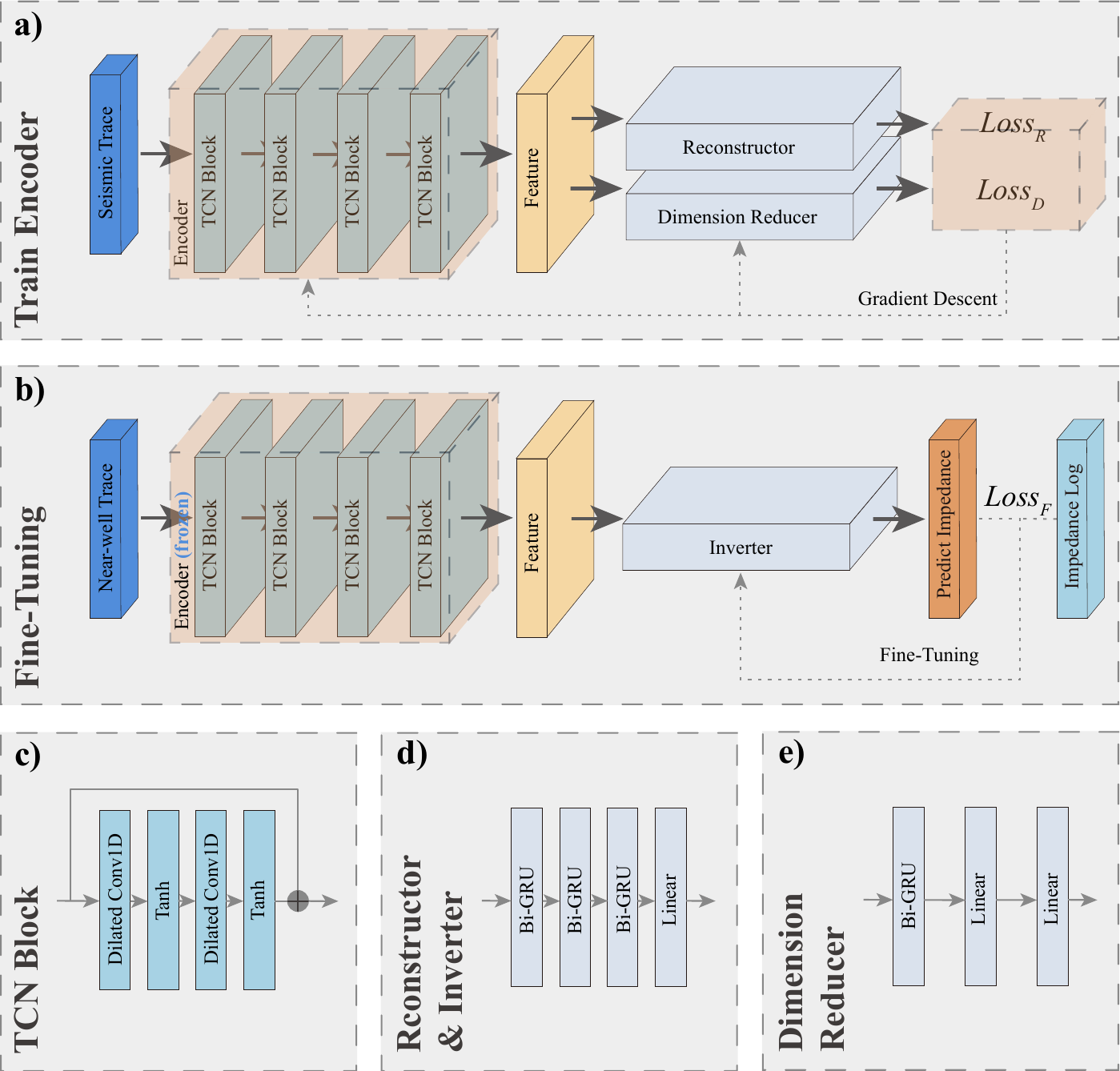}
\caption{a) The structure of the Encoder and its training strategy; b) fine-tuning of the Inverter; c), d), and e) respectively illustrate the structures of the TCN block, Inverter, Reconstructor, and Dimension Reducer.}
\label{model}
\end{figure}

The overall architecture of the proposed method and its training strategy are depicted in Figure~\ref{model}. The Encoder is constructed using four temporal convolutional network (TCN) blocks, enabling effective extraction of complex temporal features. Both the Inverter and Reconstructor adopt the same structure, each comprising bidirectional gated recurrent units (Bi-GRU) and linear layers. The Dimension Reducer, designed for adaptive feature dimensionality reduction, consists of a single Bi-GRU layer and a linear layer, resulting in fewer than 1,000 learnable parameters (depending on the seismic trace length). Overall, the proposed framework features a streamlined and efficient design.

\subsubsection{Encoder}
\begin{figure}
\centering
\noindent\includegraphics[width=\textwidth]{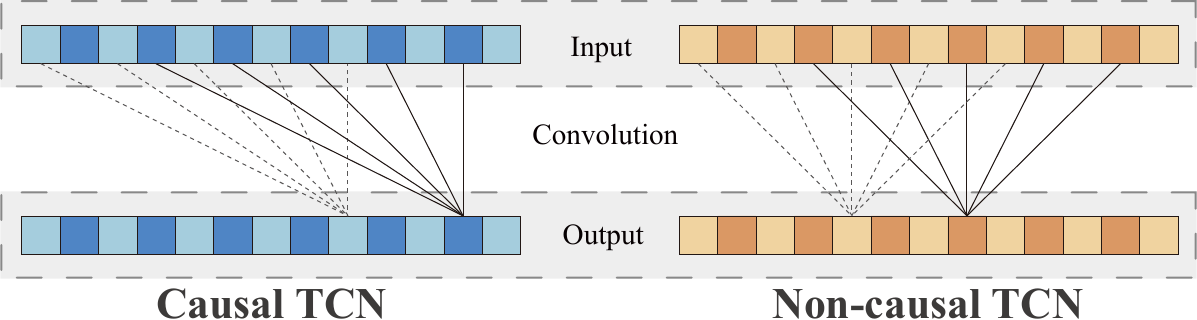}
\caption{Structures of causal and non-causal TCNs, with dilation of 2.}
\label{TCN}
\end{figure}

TCNs have demonstrated strong effectiveness in AII tasks and are widely adopted in recent studies \citep{37, 38}. TCNs are time-series models based on CNNs that leverage dilated convolutions and residual connections to capture long-range temporal dependencies. \citet{46} conducted a comprehensive analysis of various TCN architectures for inversion tasks and found that non-causal TCNs consistently yield superior performance in AII. The structures of both causal and non-causal TCNs, using a dilation of 2 as an example, are illustrated in Figure~\ref{TCN}.

The forward synthesis of seismic records, as described in Equations~\ref{Seismic Synthesis} and~\ref{Reflection Coefficient}, is inherently non-causal \citep{46}. Consequently, the AII model should also adopt a non-causal structure, which has been validated in previous studies \citep{17}. Increasing the dilation in TCNs enables the capture of long-range dependencies, but requires additional channels (kernels) to maintain sequence detail; conversely, reducing or omitting dilation necessitates deeper networks to achieve a comparable receptive field. Our objective is to map seismic traces into a high-dimensional space for linear representation. Therefore, we construct an over-complete encoder comprising four TCN blocks, as illustrated in Figure~\ref{model}c. Each TCN block contains two TCN layers with a kernel size of 3 and a dilation of 2, followed by two Tanh activation functions and a residual connection. The output feature map dimensions for the four TCN blocks are 16, 16, 16, and 32, respectively.

\subsubsection{Dimension Reducer}

As mentioned above, we hope that the features after passing through the Encoder can be as linearly separable as possible, transforming the transfer learning of the inversion model (from labeled seismic traces to unlabeled seismic traces) into linear extrapolation or interpolation. Therefore, the high-dimensional features encoded by the Encoder should be representable by high-dimensional straight lines:
\begin{equation}
F_m^x = A_{m\times 1} \times x + B_{m\times 1}, 
\end{equation}
where $F_m^x \in \{F_m^1, F_m^2,... \}$ represents the features of the consecutive seismic traces $\{ S_1(t), S_2(t),...S_x(t)\}$; $x$ represents the trace number or spatial order of the seismic traces in the entire seismic dataset. $S_x$ and $A_{m\times 1} \times x + B_{m\times 1}$ is a straight line in an m-dimensional space. However, this line is difficult to calculate directly, so we propose the Dimension Reducer for constraint. First, transform the above equation to obtain
\begin{equation}
x = \frac{A_{m\times 1}^\top }{A_{m\times 1}^\top A_{m\times 1}}(F_{m}^x-B_{m\times 1}),
\label{x}
\end{equation}
Equation~\ref{x} is in the form of a standard linear layer. Therefore, we propose the Dimension Reducer to fit this straight line, with its loss function:
\begin{equation}
Loss_D=\|D(E(S_x(t)))-x \|_2^2,
\end{equation}
where $D$ and $E$ denote the Dimension Reducer and Encoder, respectively. The architecture of the Dimension Reducer is illustrated in Figure~\ref{model}e. It consists of two linear layers (for channel and sequence length reduction) and a Bi-GRU layer with one hidden unit, which is designed to effectively constrain temporal features. As a simple and lightweight network with approximately 1,000 parameters, the Dimension Reducer provides a strong constraint on the Encoder through gradient descent. This strategy is commonly used for feature alignment in domain adaptation studies \citep{50}.

\subsubsection{Reconstructor and Inverter}

In our experiments, we observed that constraining the Encoder solely with the Dimension Reducer often led to poor convergence, as this single constraint was insufficient to ensure the effectiveness of the extracted features. To address this limitation, we introduce the Reconstructor as an additional constraint on the Encoder, as illustrated in Figure~\ref{model}d. The Reconstructor guarantees that the extracted features retain enough information to be accurately reconstructed back into seismic traces, with the following loss function:
\begin{equation}
Loss_R=\|R(E(S_x(t)))-S_x(t) \|_2^2,
\end{equation}
where $R$ represents the Reconstructor. The structure of the Reconstructor is composed of three layers of Bi-GRU with 32 hidden dimensions and a linear layer. For the Inverter, we adopt the same structure as the Reconstructor. 

On one hand, Bi-GRU combines linear layers with nonlinear activation functions. The linear layers provide strong linear extrapolation capabilities \citep{49}, while the nonlinear activation functions enable the model to capture complex, nonlinear features in acoustic impedance. On the other hand, employing different architectures for the Encoder and Reconstructor helps prevent shortcut learning and encourages the extraction of more robust and transferable features. As discussed by \citet{51}, using identical or similar structures for both components can cause the model to focus excessively on the reconstruction task, thereby hindering its generalization to inversion tasks.

Additionally, Bi-GRU offers excellent long-sequence modeling capabilities and is widely used in AII applications. Its ability to model sequences effectively without requiring very deep architectures helps mitigate overfitting in both the Reconstructor and Inverter.

\subsection{Inversion workflow}

Based on the previous sections, the AII processing workflow consists of two main stages: training the Encoder and fine-tuning the Inverter. In the first stage, the Encoder is trained using all available seismic trace records, with the Reconstructor and Dimension Reducer providing effective constraints, as illustrated in Figure~\ref{model}a and detailed in Algorithm~\ref{training encoder}. During this process, gradient descent is employed to optimize the Dimension Reducer and Reconstructor, thereby ensuring robust feature extraction by the Encoder \citep{50}.

\begin{algorithm}
        \setcounter{AlgoLine}{0}
        \scriptsize
        \LinesNumbered
        \caption{Training of the Encoder}\label{training encoder}
        \KwIn{Seismic traces $\{ S_1, S_2,...S_x\}$; Initialized Encoder $E(\theta_E)$, Reconstrustor $R(\theta_R)$, and Dimension Reducer $D(\theta_D)$}
        \KwOut{Well-trained parameters $\theta_E$}
        \;
        \;
        \For{$i=1$ to $Max\ Epoch$}{
        \For{$j=1$ to $Max\ Batch$}{
        
        Randomly extract $S_x$ from the seismic traces set 
        
        Encode the seismic trace: $F_m^x \leftarrow E(S_x,\theta_E)$ 
        
        Reduce the dimension of $F_m^x$: $x^{\prime}\leftarrow D(F_m^x, \theta_D)$ 
        
        Reconstruct the seismic trace: $S_x^{\prime}\leftarrow R(F_m^x,\theta_R)$ 
        
        Calculate the $Loss_D$: $Loss_D \leftarrow \|x-x^{\prime}\|_2^2$

        Calculate the $Loss_R$: $Loss_R \leftarrow \|S_x-S_x^{\prime}\|_2^2$

        Optimize the model parameters: $\mathop{\arg\min}\limits_{\theta_E, \theta_D, \theta_R}(Loss_R+Loss_D)$
        }
        }
        \textbf{Return:} trained parameters $\theta_E$
\end{algorithm}
Second, the trained Encoder and a small number of labeled seismic traces are used to train the Inverter, which is the fine-tuning stage and is shown in Figure~\ref{model}b and Algorithm~\ref{fine-tuning inverter}. In AII tasks, well-logging data is typically scarce, and the seismic traces with well-logging data as the true acoustic impedance account for only 1$\%$ to 5$\%$ of the total seismic traces. By applying the trained Inverter and Encoder to all seismic traces, the AII work can be completed. During the training of the Encoder, we use the Adam optimizer for gradient descent with a learning rate of 0.001. Due to the large number of seismic traces and the relatively simple encoder structure, convergence is achieved in approximately 30 epochs. In the fine-tuning stage, due to the greater optimization difficulty of AII, we use a larger learning rate (0.01) and more epochs (1,000) for training the Inverter. 

\begin{algorithm}
        \setcounter{AlgoLine}{0}
        \scriptsize
        \LinesNumbered
        \caption{Fine-tuning of the Inverter and AII}\label{fine-tuning inverter}
        \KwIn{Seismic traces $\{ S_1, S_2,...S_x\}$; Few seismic traces $\{ S_1, S_2,...S_z\}$ and matching well-logging data $\{ I_1, I_2,...I_z\}$, where $z\ll x$; Well-trained Encoder parameters $\theta_E$; Initialized Inverter $Inv(\theta_{Inv})$}
        \KwOut{Acoustic Impedance $\{ I_1, I_2,...I_x\}$}
        \tcp{fine-tuning of the Inverter}
        \;
        \For{$i=1$ to $Max\ Epoch$}{
        \For{$j=1$ to $Max\ Batch$}{
        
        Randomly extract matching $S_z$ and $I_z$ from the seismic traces and well-logging data set 
        
        Encode the seismic trace: $F_m^z \leftarrow E(S_z,\theta_E)$ 
        
        Inversion: $I_z^{\prime}\leftarrow Inv(F_m^z, \theta_{Inv})$ 

        Calculate the loss function $Loss_F$: $\|I_z-I_z^{\prime}\|_2^2$

        Optimize the model parameters: $\mathop{\arg\min}\limits_{\theta_{Inv}}Loss_F$
        }}
        \tcp{AII process}
        \For{$i=1$ to $x$}{
        $I_i \leftarrow Inv(E(S_x,\theta_E), \theta_{Inv})$
       }
        \textbf{Return:} inverted acoustic impedance $\{ I_1, I_2,...I_x\}$
\end{algorithm}
In the next section, the three most widely used open-source datasets are used for comparative experiments to verify the effectiveness and computational efficiency of the proposed method.

\section{Examples}
\subsection{Synthetic Data}
\subsubsection{Datasets and evaluation metrics}

In this section, we evaluate our method on three widely used benchmark datasets: Overthrust \citep{53}, SEAM elastic earth model \citep{54}, and Marmousi 2 \citep{52}. Each dataset is used independently for experimental validation, with no data shared between datasets. These datasets have become standard benchmarks for AII research over the past decade. The key parameters for each dataset are summarized in Table~\ref{datasets}. For all experiments, we use less than 1$\%$ of the seismic traces as labeled data, which is significantly lower than the proportions typically used in previous AII studies. Notably, the Marmousi 2 dataset presents greater structural complexity and inversion difficulty, so a slightly higher number of labeled traces is used, though still fewer than in most prior works.
\begin{table}[!htbp]
\centering
\setlength{\tabcolsep}{1.5mm}
\renewcommand\arraystretch{1.2}
\begin{tabular}{ccccc}
\toprule[1.5pt]
\textbf{Dataset} & \textbf{Trace Num}  & \textbf{Well Num} & \textbf{Time Interval} & \textbf{Record Time} \\
\midrule
Overthrust & 1067 & 10 $(\textless 1\%)$ & 4ms & 3s \\
\midrule
Marmousi 2 & 1701 & 14 $(\textless 1\%)$ & 3ms & 2.4s \\
\midrule
SEAM & 1751 & 10 $(\textless 1\%)$ & 4ms & 5s \\
\bottomrule[1.5pt]
\end{tabular}
\caption{The parameters of Overthrust, Marmousi 2 and SEAM datasets}
\label{datasets}
\end{table}

In addition, five metrics are adopted in this work for accurate comparative experiments. The first metric is the signal-to-noise ratio (SNR), which is commonly used in signal processing and is defined as
\begin{equation}
SNR=10*\log_{10}{\frac{\sum_{i=1}^{N} I_i^2}{\sum_{i=1}^{N} (I_i-I^{\prime}_i)^2}}
\label{equation: SNR}
\end{equation}
where $I$ denotes the true acoustic impedance and $I^{\prime}$ denotes the predicted acoustic impedance. SNR describes the logarithmic ratio between the true impedance and the prediction error; higher values indicate better performance. The second metric is $R^2$, the most commonly used evaluation metric in AII, defined as
\begin{equation}
R^2 = 1-\frac{\sum_{i=1}^{N}(I_i-I^{\prime}_i)^2}{\sum_{i=1}^{N}(\mu_I-I_i)^2},
\label{equation: R2}
\end{equation}
where $\mu_I$ is the mean of $I$. Structural similarity (SSIM) is a widely used metric in image processing to describe the similarity between two images, and is also used in many AII studies. Its definition is
\begin{equation}
SSIM=\frac{(2\times \mu_I \times \mu_{I^\prime} + c_1)(2 \times \sigma_{II^\prime} + c_2)}{(\mu_I^2 + \mu_{I^\prime}^2 + c_1)(\sigma_I^2 + \sigma_{I^\prime}^2 + c_2)},
\label{equation: SSIM}
\end{equation}
where $\sigma_I^2$ is the variance of $I$, $\sigma_{I^\prime}^2$ is the variance of $I^\prime$, and $\sigma_{II^\prime}$ is the covariance between $I$ and $I^\prime$. $c_1$ and $c_2$ are small positive constants to avoid division by zero. SSIM is more sensitive to global deviations and less affected by local errors. Mean absolute error (MAE) and mean squared error (MSE) are two commonly used metrics in regression tasks, defined as
\begin{equation}
MAE=\frac{1}{N}\sum_{i=1}^N | I_i-I_i^\prime|,
\label{equation: MAE}
\end{equation}
\begin{equation}
MSE=\frac{1}{N}\sum_{i=1}^N (I_i-I_i^\prime)^2.
\label{equation: MAE}
\end{equation}
Since these two metrics are affected by the data range, we standardize (Z-Score) $I$ and $I^\prime$ before calculation. 

In the following sections, we use the above metrics to conduct comparative evaluations on the datasets. Additionally, to ensure the reproducibility of our experimental results, we will open-source these datasets along with the reproducible code. 

\subsubsection{Overthrust}

\begin{figure}
\centering
\noindent\includegraphics[width=\textwidth]{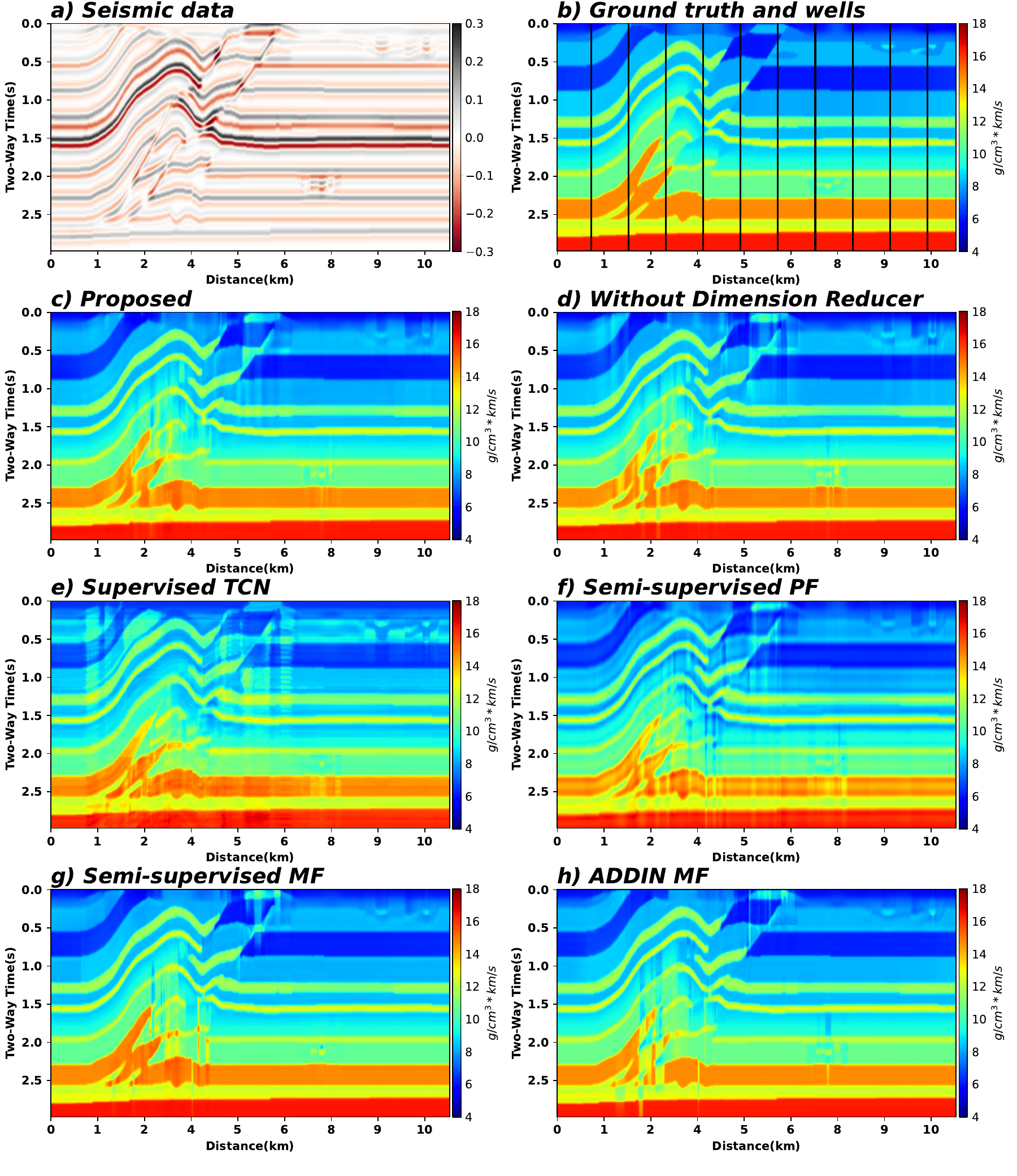}
\caption{The seismic data, true acoustic impedance and AII results of Overthrust. a) represents Overthrust seismic data; b) represents the true acoustic impedance and wells; c) to h) represent the AII results of proposed method, proposed method without Dimension Reducer, supervised TCN, semi-supervised PF, semi-supervised MF, and ADDIN MF, respectively.}
\label{overthrust_whole}
\end{figure}

The Overthrust dataset, developed by the SEG/EAGE 3-D Modeling Committee \citep{53}, is a relatively straightforward and widely adopted benchmark in AII research. For comparative analysis, we selected four open-source methods: supervised TCN \citep{19}, semi-supervised model with physics-driven forward modeling (semi-supervised PF) \citep{32}, semi-supervised model with model-driven forward modeling (semi-supervised MF) \citep{37}, and the attention-based dual-branch double-inversion network with model-driven forward modeling (ADDIN MF) \citep{38}. Additionally, we performed an ablation study to assess the contribution of the Dimension Reducer. Figures~\ref{overthrust_whole}a and b show the Overthrust seismic data and the corresponding true acoustic impedance, respectively. For AII processing, 10 seismic traces were selected at equal intervals as labeled traces, as indicated in Figure~\ref{overthrust_whole}b.
\begin{figure}
\centering
\noindent\includegraphics[width=\textwidth]{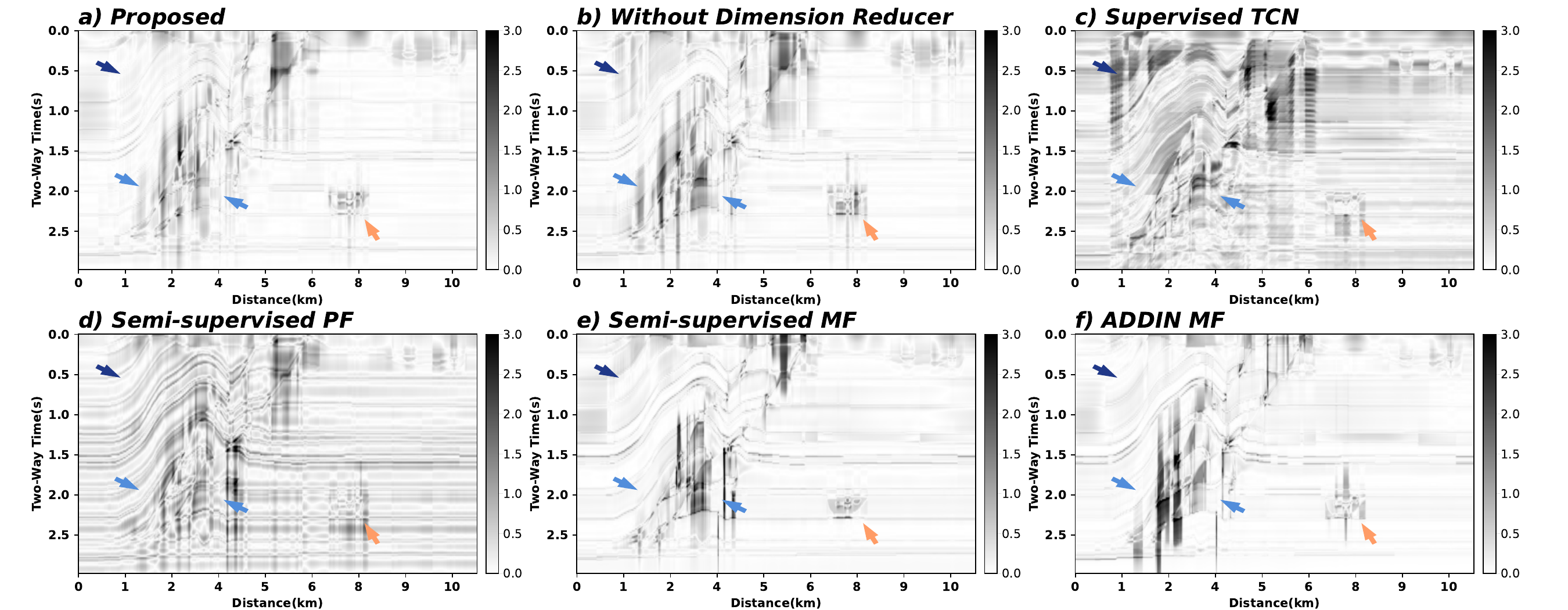}
\caption{The absolute residuals of results ($|output - groundtruth|$). a) to f) represent the absolute residuals of proposed method, proposed method without Dimension Reducer, supervised TCN, semi-supervised PF, semi-supervised MF, and ADDIN MF, respectively.}
\label{overthrust_residual}
\end{figure}
\begin{figure}
\centering
\noindent\includegraphics[width=\textwidth]{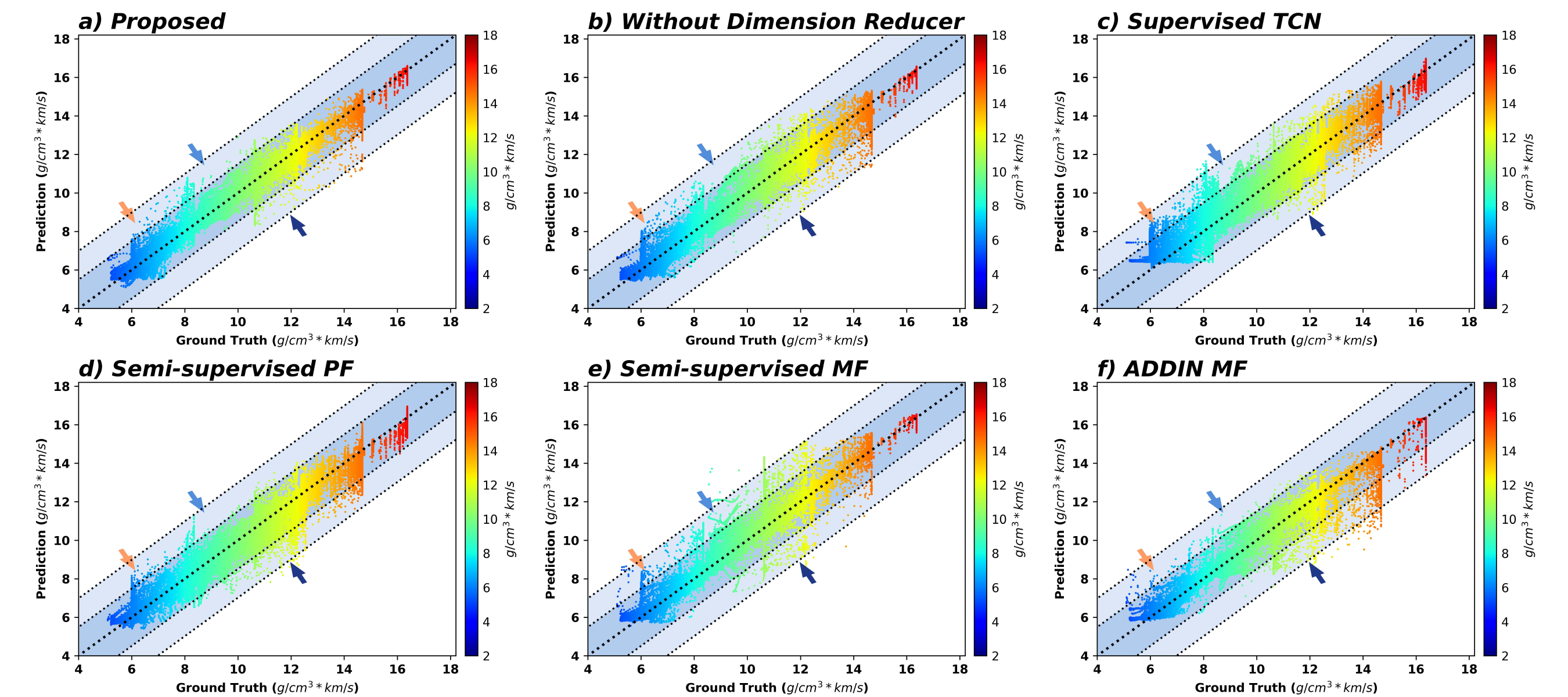}
\caption{The scatterplots of results and true acoustic impedance. a) to f) represent the scatterplots of proposed method, proposed method without Dimension Reducer, supervised TCN, semi-supervised PF, semi-supervised MF, and ADDIN MF, respectively.}
\label{overthrust_scatter}
\end{figure}
\begin{figure}
\centering
\noindent\includegraphics[width=\textwidth]{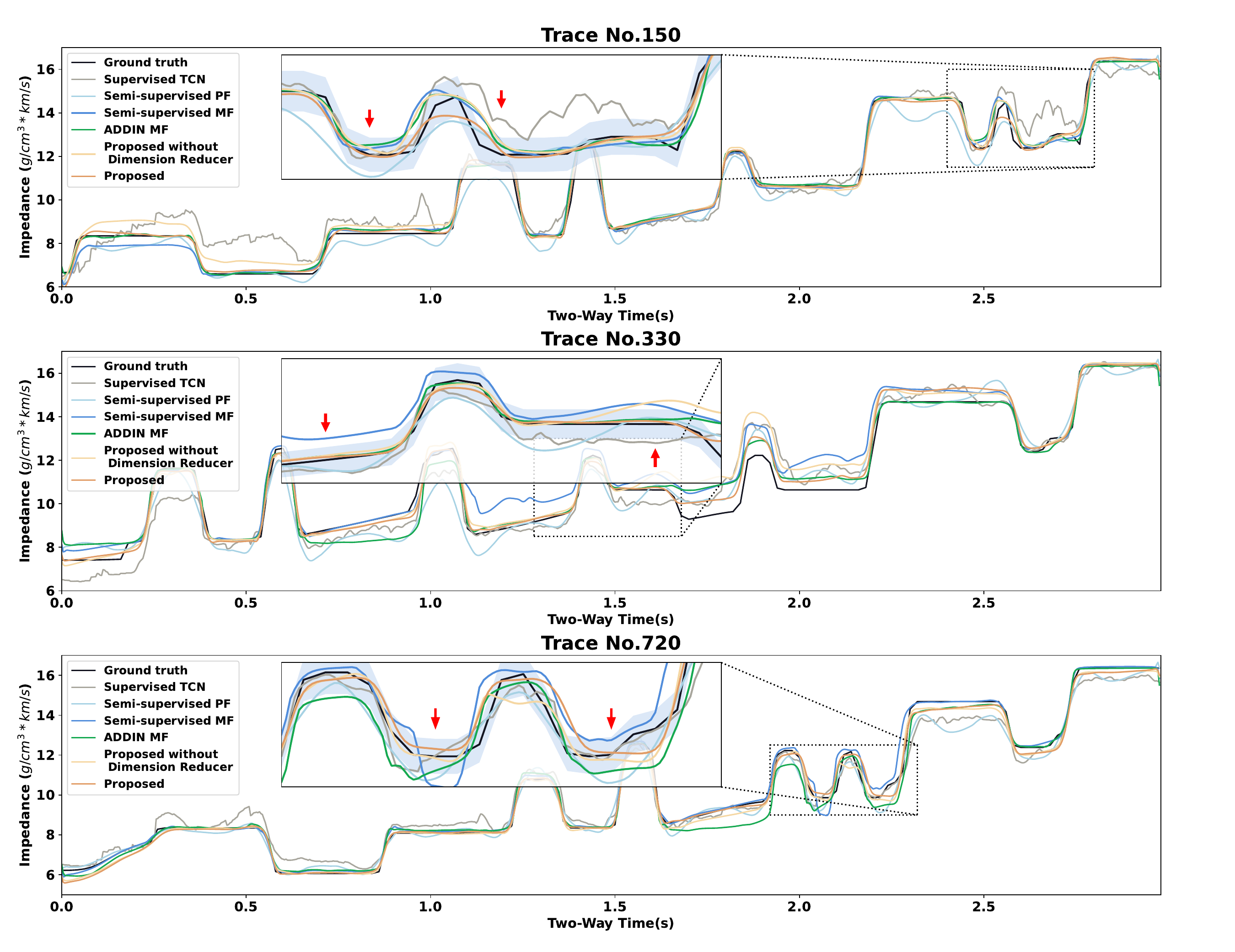}
\caption{Comparison of various methods on trace No.150, No.330, and No.720.}
\label{overthrust_trace}
\end{figure}

It can be observed that the most challenging aspect of AII arises in regions with abrupt changes, where larger deviations occur due to the scarcity of well-logging data. In particular, the supervised TCN exhibits significant errors in complex areas (Figures~\ref{overthrust_whole}e and \ref{overthrust_residual}c). By contrast, the proposed method achieves higher resolution and can clearly and accurately delineate structures such as faults (Figure~\ref{overthrust_whole}c). When the Dimension Reducer is omitted (Figure~\ref{overthrust_whole}d), lateral continuity is somewhat reduced, which becomes more pronounced in more complex datasets (Figure~\ref{overthrust_residual}b). This indicates that linear encoding of seismic traces substantially enhances the Inverter's ability to process unlabeled traces. The errors in the results of the other three methods are noticeably larger than those of the proposed method (Figure~\ref{overthrust_whole}f to h). Benefiting from a more advanced structure and attention mechanism, ADDIN MF achieves slightly better results than semi-supervised PF and semi-supervised MF (Figure~\ref{overthrust_residual}d to f).

To more intuitively illustrate the deviations among different methods, scatterplots of the predicted versus true impedance are presented in Figure~\ref{overthrust_scatter}. As shown in Figure~\ref{overthrust_scatter}a, the proposed method yields smaller deviations compared to the other approaches. Semi-supervised MF, semi-supervised PF, and ADDIN MF follow (Figures~\ref{overthrust_scatter}d to f), with only a few regions exhibiting larger deviations than the proposed method. Finally, three representative inverted traces are displayed in Figure~\ref{overthrust_trace}, further demonstrating the effectiveness of the proposed method. Overall, the Overthrust dataset is relatively straightforward; therefore, in the next section, a more complex dataset is employed for further comparison.

\subsubsection{Marmousi 2}
\begin{figure}
\centering
\noindent\includegraphics[width=\textwidth]{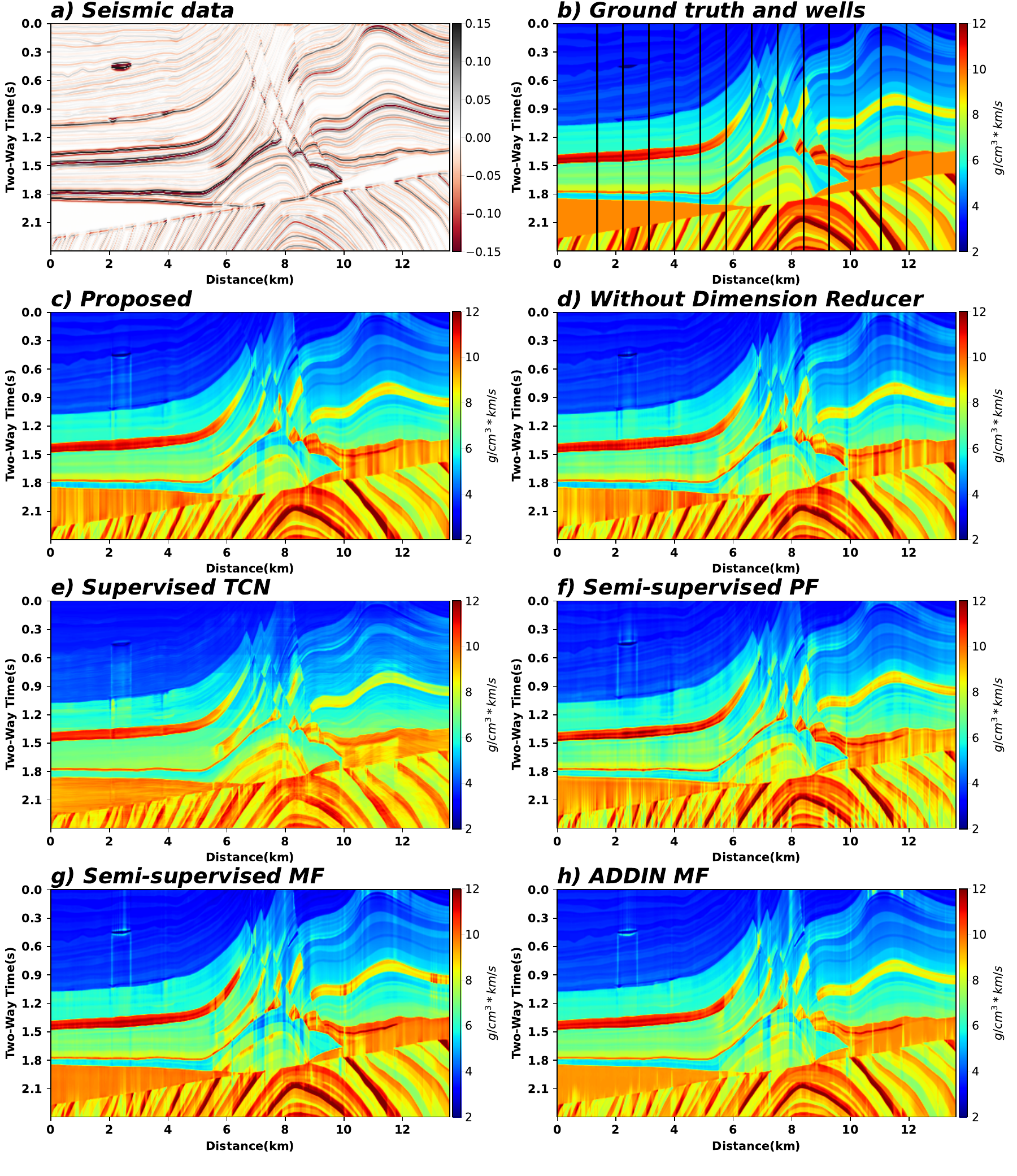}
\caption{The seismic data, true acoustic impedance and AII results of Marmousi 2. a) represents Marmousi 2 seismic data; b) represents the true acoustic impedance and wells; c) to h) represent the AII results of proposed method, proposed method without Dimension Reducer, supervised TCN, semi-supervised PF, semi-supervised MF, and ADDIN MF, respectively.}
\label{marmousi_whole}
\end{figure}

The AGL elastic Marmousi dataset, commonly referred to as Marmousi 2, was developed by \citet{52}. This dataset is highly complex, spanning 17 kilometers laterally and reaching a depth of 3.5 kilometers, and is the most widely used benchmark in the AII field. For our experiments, we utilize only 14 traces as labeled data, which is significantly fewer than the number typically used in previous studies. The seismic data and the selected labeled traces are shown in Figure~\ref{marmousi_whole}a and b.
\begin{figure}
\centering
\noindent\includegraphics[width=\textwidth]{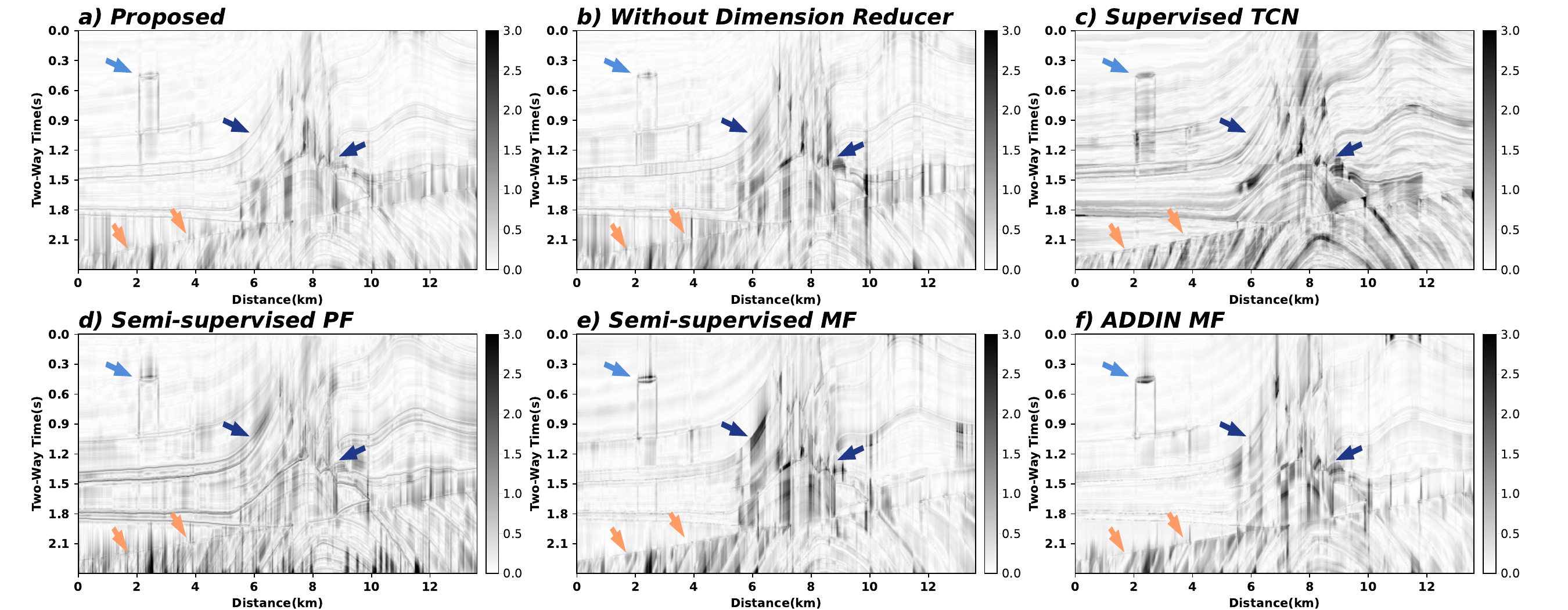}
\caption{The absolute residuals of results ($|output - groundtruth|$). a) to f) represent the absolute residuals of proposed method, proposed method without Dimension Reducer, supervised TCN, semi-supervised PF, semi-supervised MF, and ADDIN MF, respectively.}
\label{marmousi_residual}
\end{figure}
\begin{figure}
\centering
\noindent\includegraphics[width=\textwidth]{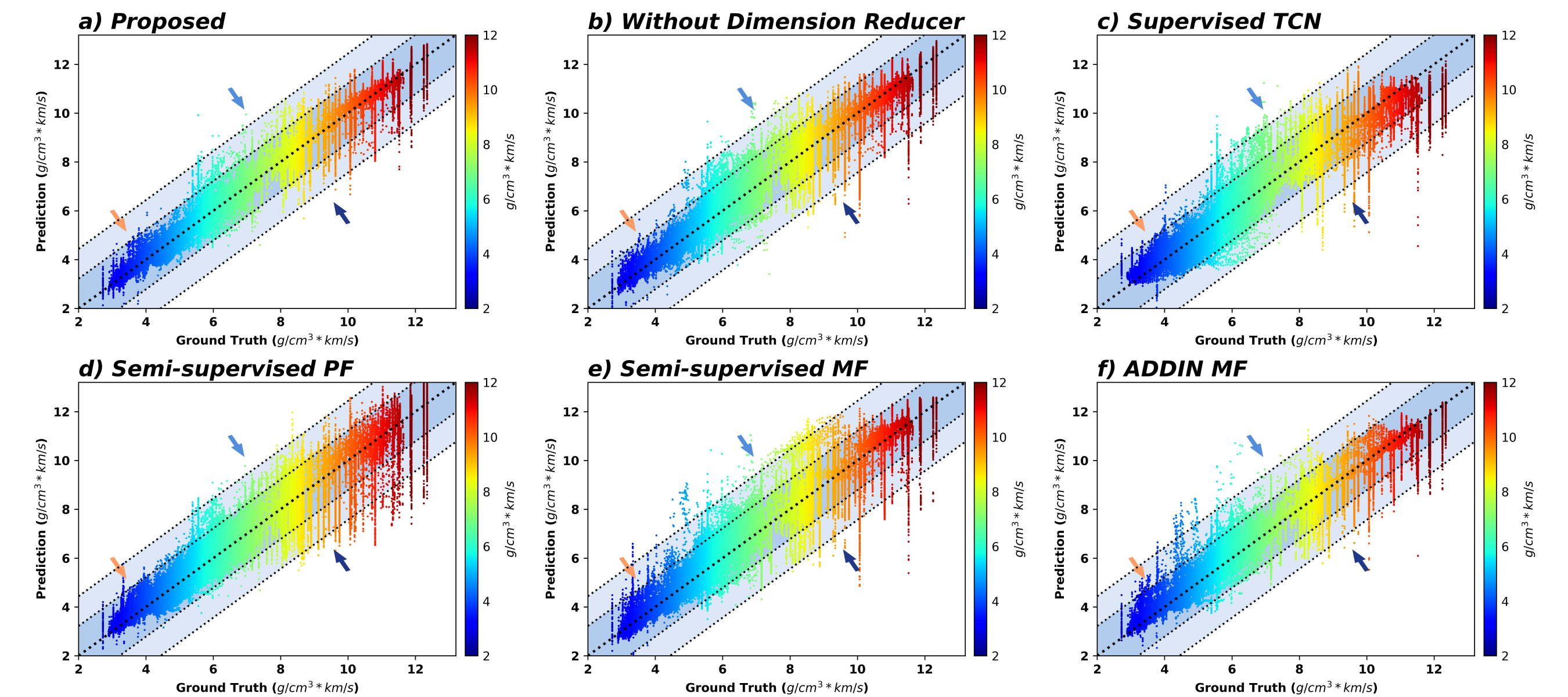}
\caption{The scatterplots of results and true acoustic impedance. a) to f) represent the scatterplots of proposed method, proposed method without Dimension Reducer, supervised TCN, semi-supervised PF, semi-supervised MF, and ADDIN MF, respectively.}
\label{marmousi_scatter}
\end{figure}
\begin{figure}
\centering
\noindent\includegraphics[width=\textwidth]{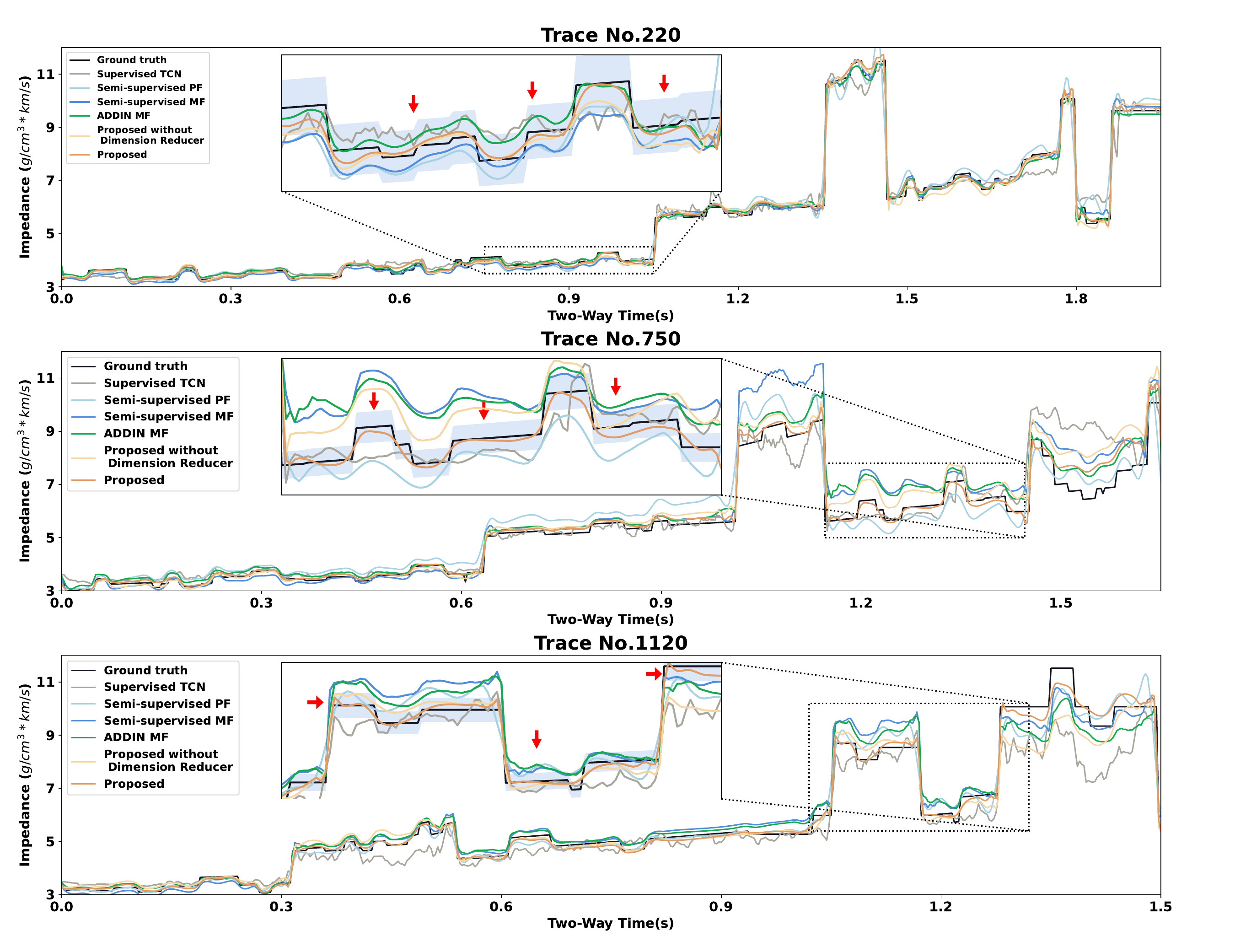}
\caption{Comparison of various methods on trace No.220, No.750, and No.1120.}
\label{marmousi_trace}
\end{figure}
Despite the increased complexity of Marmousi 2 compared to Overthrust, the proposed method consistently delivers superior results (Figure~\ref{marmousi_whole}c and Figure~\ref{marmousi_residual}a), with more pronounced improvements over competing approaches. In contrast, the supervised TCN exhibits significant overall deviations and artifacts; although it can recover the general trend of acoustic impedance, it fails to accurately capture regions with weak reflections and fine details (Figure~\ref{marmousi_whole}e and Figure~\ref{marmousi_residual}c). The differences among the semi-supervised methods are relatively modest. Semi-supervised PF shows a noticeable loss of detail in weak reflection areas (Figure~\ref{marmousi_residual}d), while semi-supervised MF and ADDIN MF each demonstrate certain advantages, but both still exhibit considerable errors in complex regions (Figure~\ref{marmousi_residual}e and f). These discrepancies are further illustrated by the scatterplots in Figure~\ref{marmousi_scatter}.

As shown in Figures~\ref{marmousi_scatter}a and b, the proposed Dimension Reducer effectively constrains the encoder as intended. With the linear constraint on features, the overall error of AII is substantially reduced. Additionally, several representative inverted seismic traces are presented in Figure~\ref{marmousi_trace}, highlighting the complexity of the Marmousi 2 model and the more evident differences among various methods.

\subsubsection{SEAM}
\begin{figure}
\centering
\noindent\includegraphics[width=\textwidth]{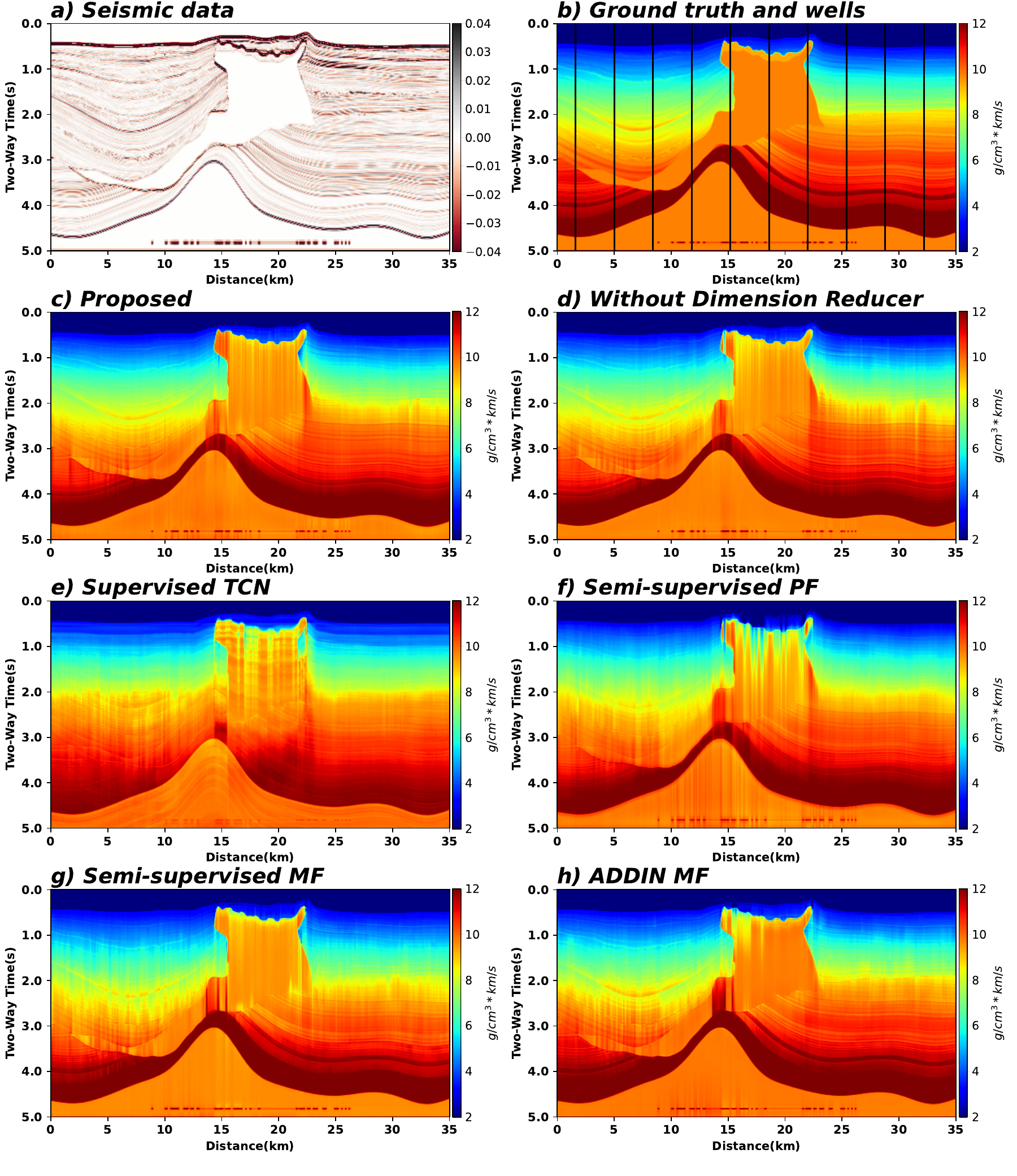}
\caption{The seismic data, true acoustic impedance and AII results of SEAM. a) represents SEAM seismic data; b) represents the true acoustic impedance and wells; c) to h) represent the AII results of proposed method, proposed method without Dimension Reducer, supervised TCN, semi-supervised PF, semi-supervised MF, and ADDIN MF, respectively.}
\label{seam_whole}
\end{figure}

The SEAM elastic earth model includes a 35km survey line, and all its properties are derived from basic rock properties \citep{54}. In addition to containing a sandstone body, SEAM has a very rich number of thin layers (Figure~\ref{seam_whole}a), which is conducive to evaluating the accuracy of AII methods. 
\begin{figure}
\centering
\noindent\includegraphics[width=\textwidth]{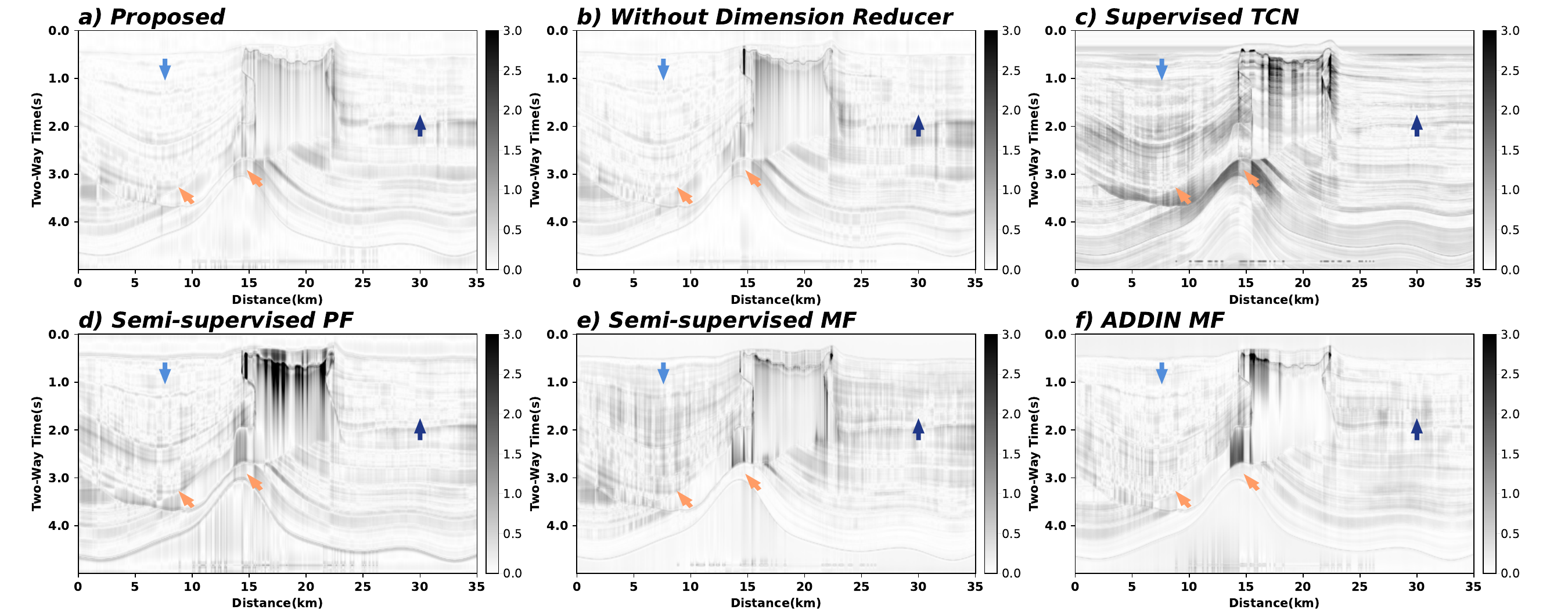}
\caption{The absolute residuals of results ($|output - groundtruth|$). a) to f) represent the absolute residuals of proposed method, proposed method without Dimension Reducer, supervised TCN, semi-supervised PF, semi-supervised MF, and ADDIN MF, respectively.}
\label{seam_residual}
\end{figure}
\begin{figure}
\centering
\noindent\includegraphics[width=\textwidth]{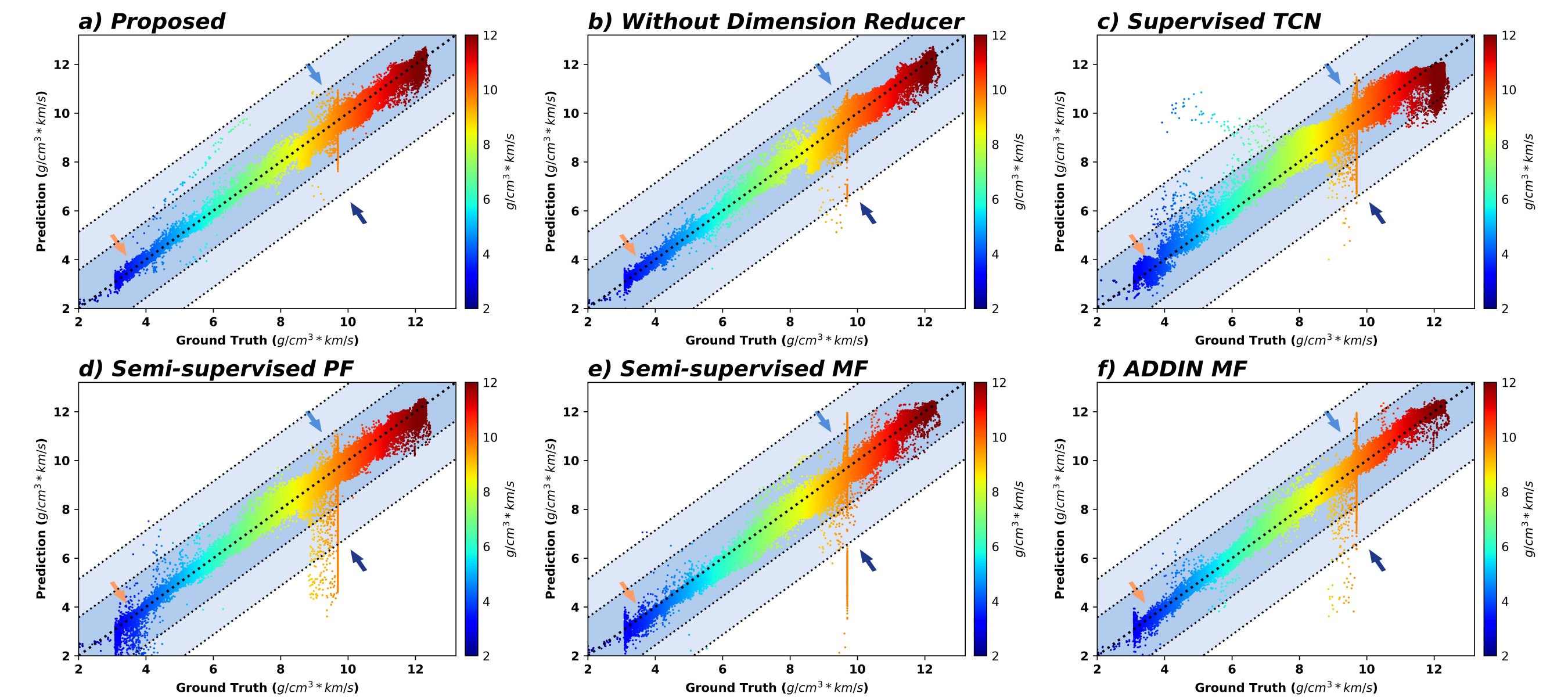}
\caption{The scatterplots of results and true acoustic impedance. a) to f) represent the scatterplots of proposed method, proposed method without Dimension Reducer, supervised TCN, semi-supervised PF, semi-supervised MF, and ADDIN MF, respectively.}
\label{seam_scatter}
\end{figure}
\begin{figure}
\centering
\noindent\includegraphics[width=\textwidth]{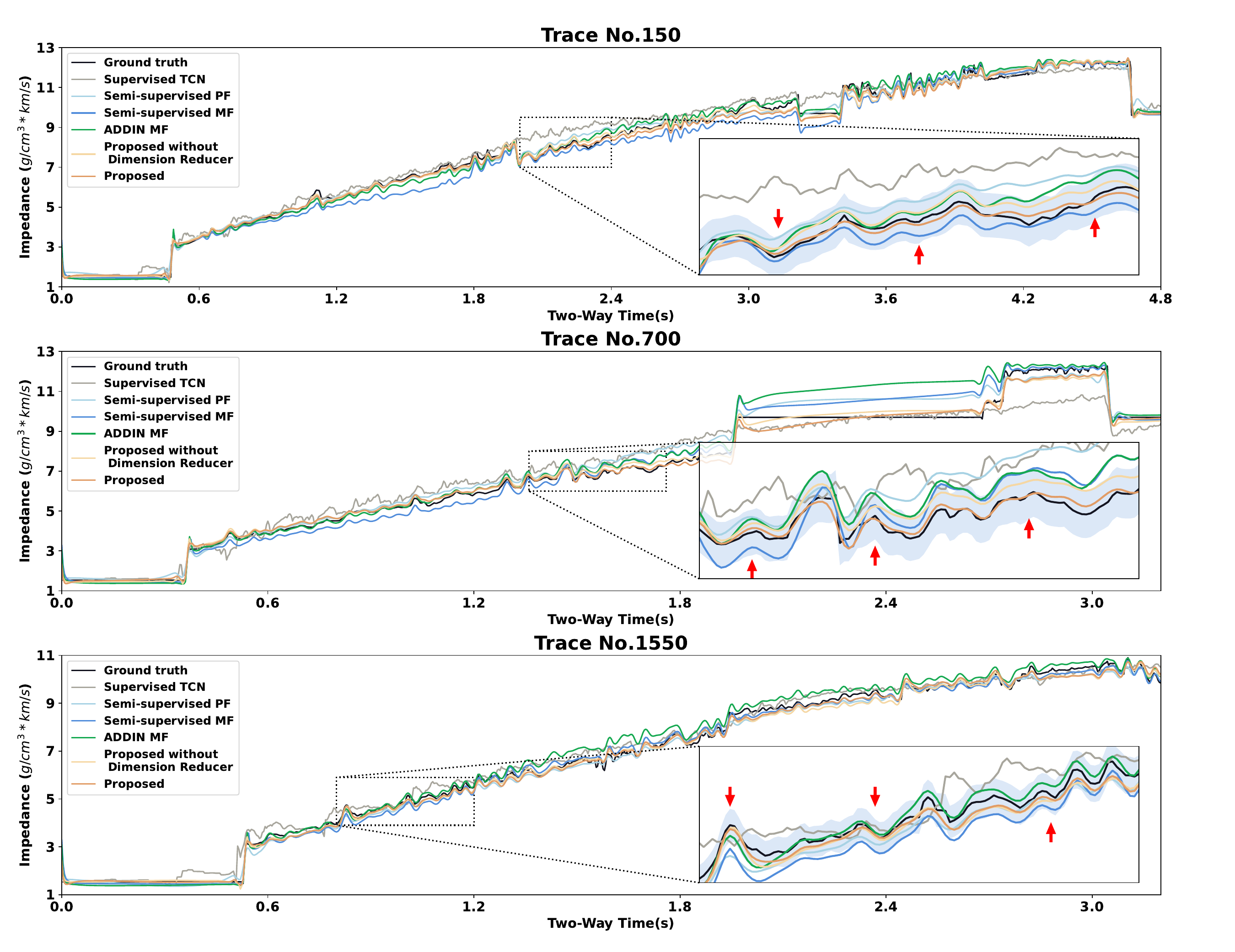}
\caption{Comparison of various methods on trace No.150, No.700, and No.1550.}
\label{seam_trace}
\end{figure}

It can be observed that the supervised TCN model almost completely loses detailed features, such as thin layers and the sandstone body (Figure~\ref{seam_whole}e and Figure~\ref{seam_residual}c). The semi-supervised PF method shows some improvement in thin-layer regions compared to the supervised TCN (Figure~\ref{seam_whole}f and Figure~\ref{seam_residual}d), but still exhibits significant errors in the sandstone body areas. In contrast, the semi-supervised MF and ADDIN MF methods demonstrate notable improvements over semi-supervised PF (Figure~\ref{seam_residual}e and f). The proposed method achieves the best results, with overall errors smaller than those of other methods in both thin-layer and sandstone body regions (Figure~\ref{seam_residual}a). As shown in Figure~\ref{seam_scatter}, except for minor deviations in a few regions, the proposed method generally maintains relatively small errors and outperforms the other comparative approaches.

Finally, Figure~\ref{seam_trace} presents three representative inverted seismic traces. Compared with the previous two datasets, the SEAM dataset contains more thin layers and weak reflection regions, increasing the difficulty of AII. In this case, the errors of the supervised TCN are pronounced, while the proposed method achieves significantly smaller errors and provides results closer to the true acoustic impedance than the semi-supervised methods.

\subsubsection{Quantitative results}

Table~\ref{results} summarizes the quantitative evaluation of all methods using five metrics across the three benchmark datasets. The proposed method consistently achieves superior performance on all metrics. Notably, when the Dimension Reducer is omitted, the performance of the proposed method becomes comparable to that of the semi-supervised MF and ADDIN MF approaches, underscoring the critical role of the linear feature constraint. These findings demonstrate that linear feature encoding substantially enhances the Inverter's ability to process unlabeled traces, thereby improving overall AII accuracy and robustness. Furthermore, while supervised methods may yield reasonable results on relatively simple datasets, their effectiveness declines markedly on more complex datasets with limited well-logging data.
\begin{table}[!htbp]
\centering
\tiny
\setlength{\tabcolsep}{1.2mm}
\renewcommand\arraystretch{1.6}
\begin{tabular}{cccccccc}
\toprule[1.5pt]
\textbf{Metric} & \textbf{Dataset}  & \textbf{Proposed} & \textbf{\makecell{Proposed \\ (without Dimension \\ Reducer)}}  & \textbf{\makecell{Supervised \\ TCN}} & \textbf{\makecell{Semi-supervised \\ PF}} & \textbf{\makecell{Semi-supervised \\ MF}} & \textbf{ADDIN MF} \\
\midrule
\multirow{3}{*}{\textbf{SNR $\uparrow $}} & Overthrust & \textbf{30.8307} & 29.5747 & 25.3310 & 27.0136 & 29.1915 & 28.9334 \\
& Marmousi 2 & \textbf{25.7436} & 23.9952 & 21.7474 & 22.3415 & 23.3248 & 24.3346 \\
& SEAM & \textbf{31.5077} & 30.3421 & 25.0340 & 25.3898 & 28.6047 & 29.0701 \\
\midrule
\multirow{3}{*}{\textbf{$R^2 \uparrow$}} & Overthrust & \textbf{0.9892} & 0.9856 & 0.9616 & 0.9740 & 0.9842 & 0.9833 \\
& Marmousi 2 & \textbf{0.9783} & 0.9675 & 0.9455 & 0.9524 & 0.9621 & 0.9699 \\
& SEAM & \textbf{0.9944} & 0.9926 & 0.9749 & 0.9769 & 0.9890 & 0.9901 \\
\midrule
\multirow{3}{*}{\textbf{SSIM $\uparrow $}} & Overthrust & \textbf{0.9542} & 0.9330 & 0.8587 & 0.8809 & 0.9359 & 0.9111 \\
& Marmousi 2 & \textbf{0.9134} & 0.8907 & 0.8108 & 0.8355 & 0.8827 & 0.8948 \\
& SEAM & \textbf{0.9277} & 0.9131 & 0.8296 & 0.8711 & 0.9245 & 0.9124 \\
\midrule
\multirow{3}{*}{\textbf{MAE $\downarrow$}} & Overthrust & \textbf{0.0579} & 0.0674 & 0.1346 & 0.1126 & 0.0627 & 0.0636\\
& Marmousi 2 & \textbf{0.0914} & 0.1128 & 0.1602 & 0.1444 & 0.1130 & 0.0988 \\
& SEAM & \textbf{0.0497} & 0.0547 & 0.1086 & 0.0851 & 0.0583 & 0.0510 \\
\midrule
\multirow{3}{*}{\textbf{MSE $\downarrow$}} & Overthrust & \textbf{0.0108} & 0.0145 & 0.0351 & 0.0244 & 0.0153 & 0.0166 \\
& Marmousi 2 & \textbf{0.0217} & 0.0327 & 0.0547 & 0.0480 & 0.0369 & 0.0290 \\
& SEAM & \textbf{0.0055} & 0.0072 & 0.0243 & 0.0227 & 0.0088 & 0.0082 \\
\bottomrule[1.5pt]
\end{tabular}
\caption{The quantitative metrics of various methods on Overthrust, Marmousi 2 and SEAM datasets.}
\label{results}
\end{table}

There are also notable differences among the three semi-supervised methods. The semi-supervised approach based on physics-driven forward modeling relies on accurate seismic wavelet extraction and is highly sensitive to wavelet quality. In contrast, the semi-supervised method utilizing a deep learning model-driven forward process alleviates this limitation; however, the reliability of data-driven constraints remains challenging, indicating potential for further improvement. Compared to these approaches, the proposed method demonstrates superior robustness and AII performance, achieving substantial improvements in both accuracy and precision.

\subsection{Field Data}

\begin{figure}
\centering
\noindent\includegraphics[width=\textwidth]{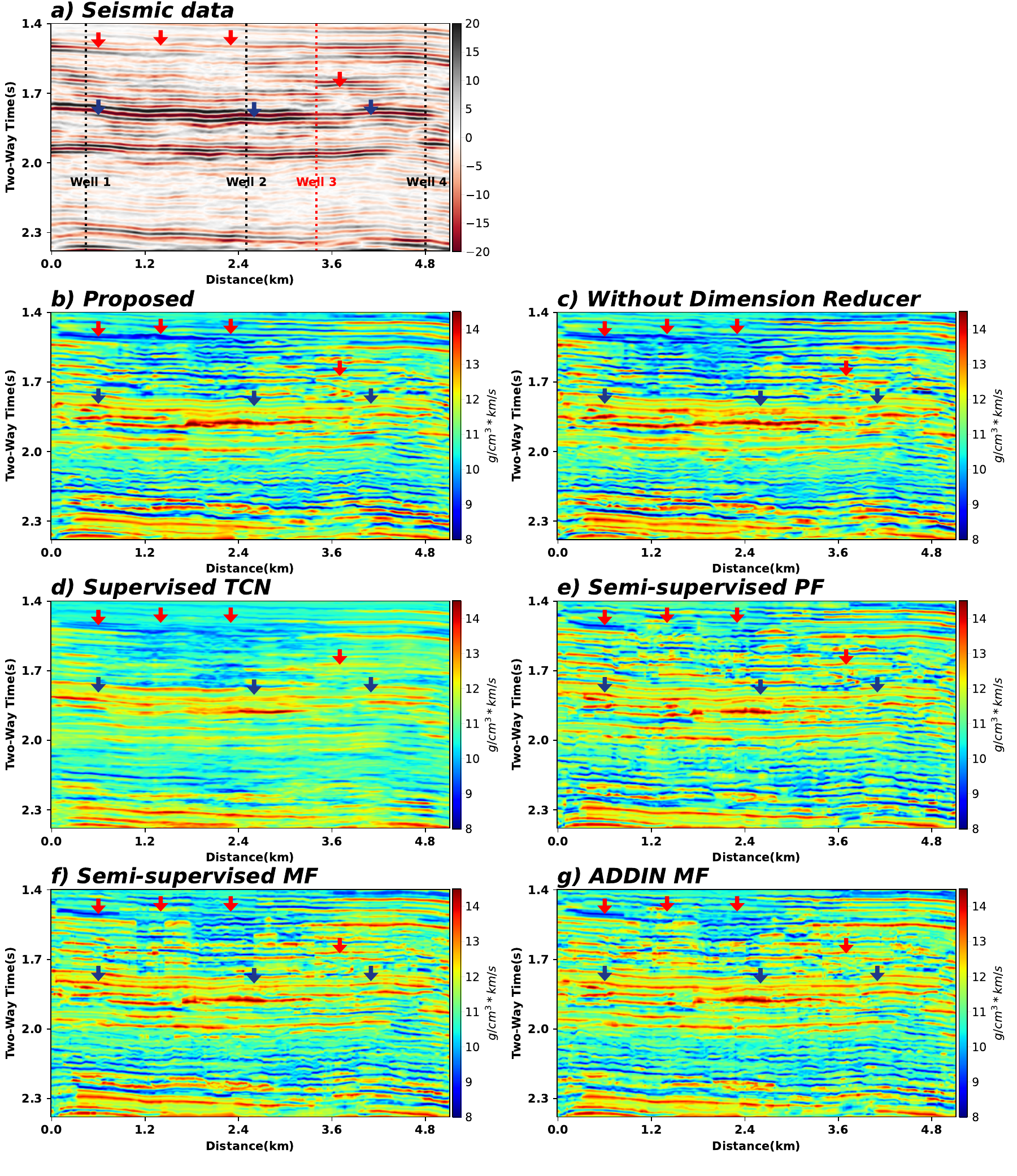}
\caption{The seismic data, wells and AII results of field data. a) represents field seismic data and wells; b) to g) represent the AII results of proposed method, proposed method without Dimension Reducer, supervised TCN, semi-supervised PF, semi-supervised MF, and ADDIN MF, respectively.}
\label{field_whole}
\end{figure}

In addition to synthetic data, we further evaluate our method using a field dataset from the Sichuan Basin, a representative continental sedimentary basin in southwestern China, as shown in Figure~\ref{field_whole}. The Sichuan Basin is characterized by multi-phase tectonic activity, resulting in complex geological structures, pronounced lateral lithological variations, and significant reservoir heterogeneity. The primary target strata consist of interbedded sandstones and mudstones, where tight gas reservoirs exhibit low porosity, low permeability, and strong heterogeneity, with frequent lateral and vertical lithofacies changes. Distinct impedance contrasts between high-impedance sandstones and low-impedance mudstones are critical for seismic acoustic impedance inversion. Additionally, the region has experienced multiple tectonic events, leading to the development of faults and fractures, which further complicate seismic responses and increase the difficulty of inversion tasks. The field dataset contains four wells; three wells (Well 1, 2, and 4) are used for training, and one well (Well 3) is reserved for validation. All methods are able to accurately invert the high-impedance surrounding layers (as indicated by the blue arrows), and the high-impedance layers enclosing low-impedance layers are consistent with the characteristics of tight gas reservoirs and the regional geological background.

\begin{figure}
\centering
\noindent\includegraphics[width=\textwidth]{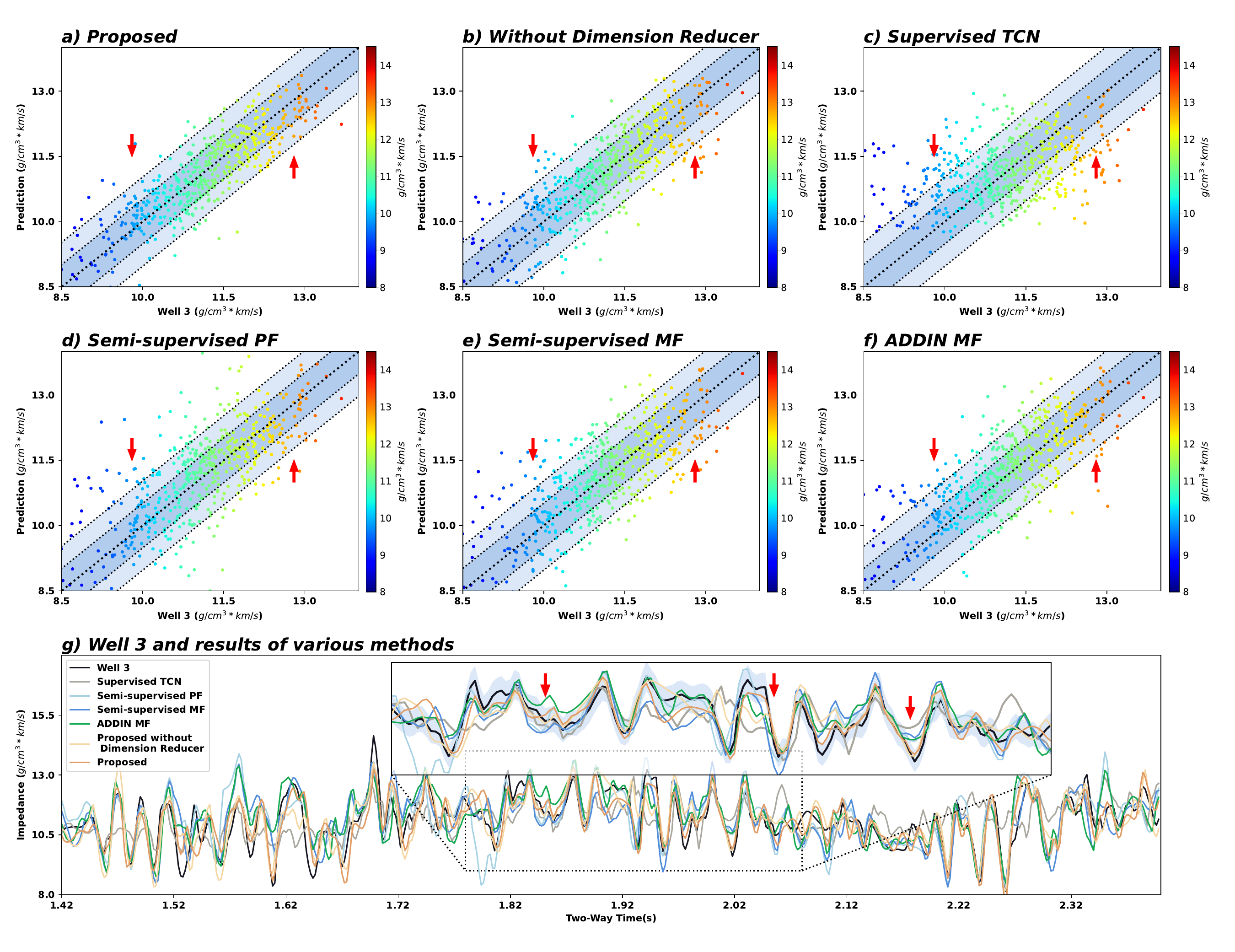}
\caption{The scatterplots and curves of results and Well 3. a) to f) represent the Well 3 scatterplots of proposed method, proposed method without Dimension Reducer, supervised TCN, semi-supervised PF, semi-supervised MF, and ADDIN MF, respectively; g) represents comparison of various methods on Well 3.}
\label{field_scatter_trace}
\end{figure}

For further development, achieving high-resolution inversion results is essential. However, the supervised TCN method exhibits significant errors in fine details, making it less effective than other inversion approaches. Additionally, all three supervised methods lack lateral continuity compared to the proposed method (as indicated by the red arrows) and show a noticeable loss of thin layers. We also present scatter plots comparing the results of various methods with those from Well 3. As shown in Figure~\ref{field_scatter_trace}a to f, all methods display some degree of deviation, which is expected given that field well-log data may contain acquisition and processing errors. Nevertheless, the proposed method demonstrates deviations within a reasonable range (as indicated by the red arrows). Figure~\ref{field_scatter_trace}g further compares the methods, with $R^2$ values for Well 3 as follows: 0.7354 (proposed method), 0.6301 (proposed method without Dimension Reducer), 0.1704 (supervised TCN), 0.3918 (semi-supervised PF), 0.4912 (semi-supervised MF), and 0.5107 (ADDIN MF). Although all methods generally follow the trend of Well 3, the proposed method achieves better detail preservation and more accurately inverts the main horizons, resulting in overall closer agreement with Well 3.

While the proposed method demonstrates promising results on field data, certain limitations remain. First, the quantity and quality of well-log data in field applications are often limited, and the available data may not fully capture the lateral and vertical heterogeneity of the subsurface. This can reduce inversion accuracy, particularly in areas far from well control or with complex geological structures. Second, the reference acoustic impedance in field settings is typically derived from well logs, which may be affected by measurement errors, depth mismatches, or incomplete coverage. These factors further complicate the objective evaluation of inversion performance.

\section{Discussion}
\subsection{Robustness and Efficiency}

In AII, the noise level of seismic data, wavelet characteristics, and the number of available wells can vary considerably across different exploration scenarios. To evaluate the robustness of the proposed method, we conducted experiments assessing its noise tolerance, wavelet generalization capability, and sensitivity to the number of wells (see Table~\ref{noise-wavelet-well}). As expected, AII performance improves with increasing seismic data SNR, and the method remains effective even under moderate noise conditions. When the SNR reaches 15 or higher, inversion results are comparable to those obtained with noise-free data. However, seismic data with SNR below 5 are typically considered highly noisy and generally require noise attenuation preprocessing \citep{12}. 

Wavelet characteristics also play a critical role in AII. In addition to the commonly used Ricker wavelet, Berlage \citep{58} and generalized \citep{59} wavelets are frequently employed. The Berlage wavelet, a minimum-phase wavelet with a simple structure and more pronounced wave crests, yields noticeably improved AII results compared to the other types. The generalized wavelet, as an extension of the Ricker wavelet, produces inversion results similar to those of the Ricker wavelet. In field applications, seismic wavelets are often more complex. In future work, we plan to address this by incorporating wavelet variability into the training process to further enhance model performance.

\begin{table}[!htbp]
\centering
\tiny
\setlength{\tabcolsep}{1.2mm}
\renewcommand\arraystretch{1.6}
\begin{tabular}{ccccccccccc}
\toprule[1.5pt]
\textbf{Metric} & \textbf{Dataset} & \multicolumn{3}{c}{\textbf{SNR of Seismic Data}} & \multicolumn{3}{c}{\textbf{Wavelet Types}} & \multicolumn{3}{c}{\textbf{Well Nums}}\\
\cline{3-5}\cline{6-8}\cline{9-11}
 & & \textbf{5} & \textbf{10} & \textbf{15} & Ricker & Berlage & Generalized & 4 & 6 & 8 \\
\midrule
\multirow{3}{*}{\textbf{SNR $\uparrow $}} & Overthrust & 25.9874 & 29.3279 & 30.1879 & 30.8307 & 35.0346 & 29.7337 & 26.9841 & 28.2030 & 29.8369\\
& Marmousi 2 & 21.5165 & 23.0341 & 24.9827 & 25.7436 & 28.1581 & 24.6795 & 20.3520 & 21.5334 & 23.4407\\
& SEAM & 28.6758 & 30.5803 & 31.4851 & 31.5077 & 36.0545 & 32.3438 & 24.4786 & 30.2454 & 30.6746\\
\midrule
\multirow{3}{*}{\textbf{$R^2 \uparrow$}} & Overthrust & 0.9670 & 0.9847 & 0.9878 & 0.9892 & 0.9959 & 0.9861 & 0.9738 & 0.9802 & 0.9864 \\
& Marmousi 2 & 0.9425 & 0.9594 & 0.9741 & 0.9783 & 0.9875 & 0.9722 & 0.9248 & 0.9427 & 0.9631\\
& SEAM & 0.9892 & 0.9930 & 0.9943 & 0.9944 & 0.9980 & 0.9953 & 0.9715 & 0.9924 & 0.9932\\
\midrule
\multirow{3}{*}{\textbf{SSIM $\uparrow $}} & Overthrust & 0.6998 & 0.8480 & 0.9087 & 0.9542 & 0.9738 & 0.9524 & 0.9371 & 0.9446 & 0.9468\\
& Marmousi 2 & 0.6582 & 0.7843 & 0.8622 & 0.9134 & 0.9545 & 0.9131 & 0.8864 & 0.8502 & 0.8881\\
& SEAM & 0.7859 & 0.8696 & 0.9001 & 0.9277 & 0.9252 & 0.9131 & 0.8846 & 0.9027 & 0.9143\\
\midrule
\multirow{3}{*}{\textbf{MAE $\downarrow$}} & Overthrust & 0.1184 & 0.0771 & 0.0619 & 0.0579 & 0.0396 & 0.0593 & 0.0938 & 0.0821 & 0.0656\\
& Marmousi 2 & 0.1629 & 0.1320 & 0.1034 & 0.0914 & 0.0683 & 0.0992 & 0.1808 & 0.1694 & 0.1233\\
& SEAM & 0.0722 & 0.0579 & 0.0521 & 0.0497 & 0.0281 & 0.0456 & 0.0774 & 0.0556 & 0.0529\\
\midrule
\multirow{3}{*}{\textbf{MSE $\downarrow$}} & Overthrust & 0.0330 & 0.0153 & 0.0102 & 0.0101 & 0.0041 & 0.0137 & 0.0262 & 0.0198 & 0.0136\\
& Marmousi 2 & 0.0577 & 0.0408 & 0.0260 & 0.0217 & 0.0124 & 0.0275 & 0.0692 & 0.0575 & 0.0372\\
& SEAM & 0.0108 & 0.0070 & 0.0057 & 0.0055 & 0.0020 & 0.0137 & 0.0283 & 0.0074 & 0.0067\\
\bottomrule[1.5pt]
\end{tabular}
\caption{The results of the proposed method on each dataset under different seismic data noise levels, different wavelet types, and different numbers of wells.}
\label{noise-wavelet-well}
\end{table}

From a cost perspective, drilling a single well is extremely expensive, making it essential to address AII in scenarios with limited well availability. As shown in Table~\ref{noise-wavelet-well}, the proposed method achieves satisfactory results with as few as 8 wells, although performance decreases slightly as the number of wells is reduced. It should be noted that the required number of wells depends on subsurface complexity; for example, more complex datasets such as Marmousi 2 require additional wells to achieve higher AII accuracy. Effectively performing AII with a small number of wells remains one of the most challenging problems in the field.

Efficiency is also a key consideration for the proposed method. Table~\ref{parameter-efficiency} summarizes the number of parameters for each method and the time required to process each dataset (using 10 wells for Overthrust and SEAM, and 14 wells for Marmousi 2) as well as the field data. All methods were evaluated under consistent conditions: Intel Core i5-14400F processor, 16GB memory, NVIDIA RTX 4060 GPU (6GB), Python 3.8.20, and PyTorch 2.4.1, with a fixed random seed for initialization. Although the total number of parameters in all components of the proposed method is relatively large, at most approximately 57.5k parameters are involved during Encoder or Inverter training. Moreover, since our method does not require a forward modeling process, the total number of training iterations is lower than that of other methods, resulting in reduced overall computation time. The supervised TCN requires more iterations for complex data \citep{19}, while semi-supervised methods require even more iterations due to learning from both labeled and unlabeled seismic traces, further increasing training time. Model efficiency is also influenced by network architecture; for example, although the semi-supervised MF method has more parameters than ADDIN MF, its overall computation time is lower because ADDIN MF employs an attention mechanism, which, despite having relatively few parameters, is more computationally intensive than standard convolutional or fully connected layers. Therefore, further improvements in network design, reduction of learnable parameters, and enhancement of computational efficiency are important directions for future research.

\begin{table}[!htbp]
\centering
\tiny
\setlength{\tabcolsep}{1.2mm}
\renewcommand\arraystretch{1.6}
\begin{tabular}{ccccccc}
\toprule[1.5pt]
 \multicolumn{2}{c}{\textbf{Metric}} & \textbf{Proposed Methods}  & \textbf{\makecell{Supervised \\ TCN}} & \textbf{\makecell{Semi-supervised \\ PF}} & \textbf{\makecell{Semi-supervised \\ MF}} & \textbf{ADDIN MF}\\
\midrule
 \multicolumn{2}{c}{\textbf{Parameter Num}} & \makecell{Encoder: 6128\\ Inverter: 50369 \\ Reconstructor: 50369 \\ Dimension Reducer: 1000} & 2026 & 7297 & \makecell{Inverse: 138595\\ Forward: 48} & \makecell{Inverse: 32962 \\ Forward: 8738} \\
\midrule
\multirow{4}{*}{\textbf{Time(s) $\downarrow$}} & \textbf{Overthrust} & \textbf{30} & 32 & 58 & 69 & 71 \\
& \textbf{Marmousi 2} & 51 & \textbf{34} & 112 & 136 & 141 \\
& \textbf{SEAM} & 68 & \textbf{35} & 125 & 153 & 157 \\
& \textbf{Field data} & \textbf{9} & 31 & 34 & 38 & 51 \\
\bottomrule[1.5pt]
\end{tabular}
\caption{The numbers of parameter and time consumption of each method.}
\label{parameter-efficiency}
\end{table}

\subsection{Dimension Reducer and Reconstructor}

In this work, we propose linear feature encoding of seismic data to enhance the accuracy of acoustic impedance inversion (AII). Drawing inspiration from domain adversarial learning \citep{50}, we utilize the Dimension Reducer and gradient descent to constrain the Encoder. Experimental results demonstrate the effectiveness of linear feature encoding, with several notable advantages.

\begin{figure}
\centering
\noindent\includegraphics[width=\textwidth]{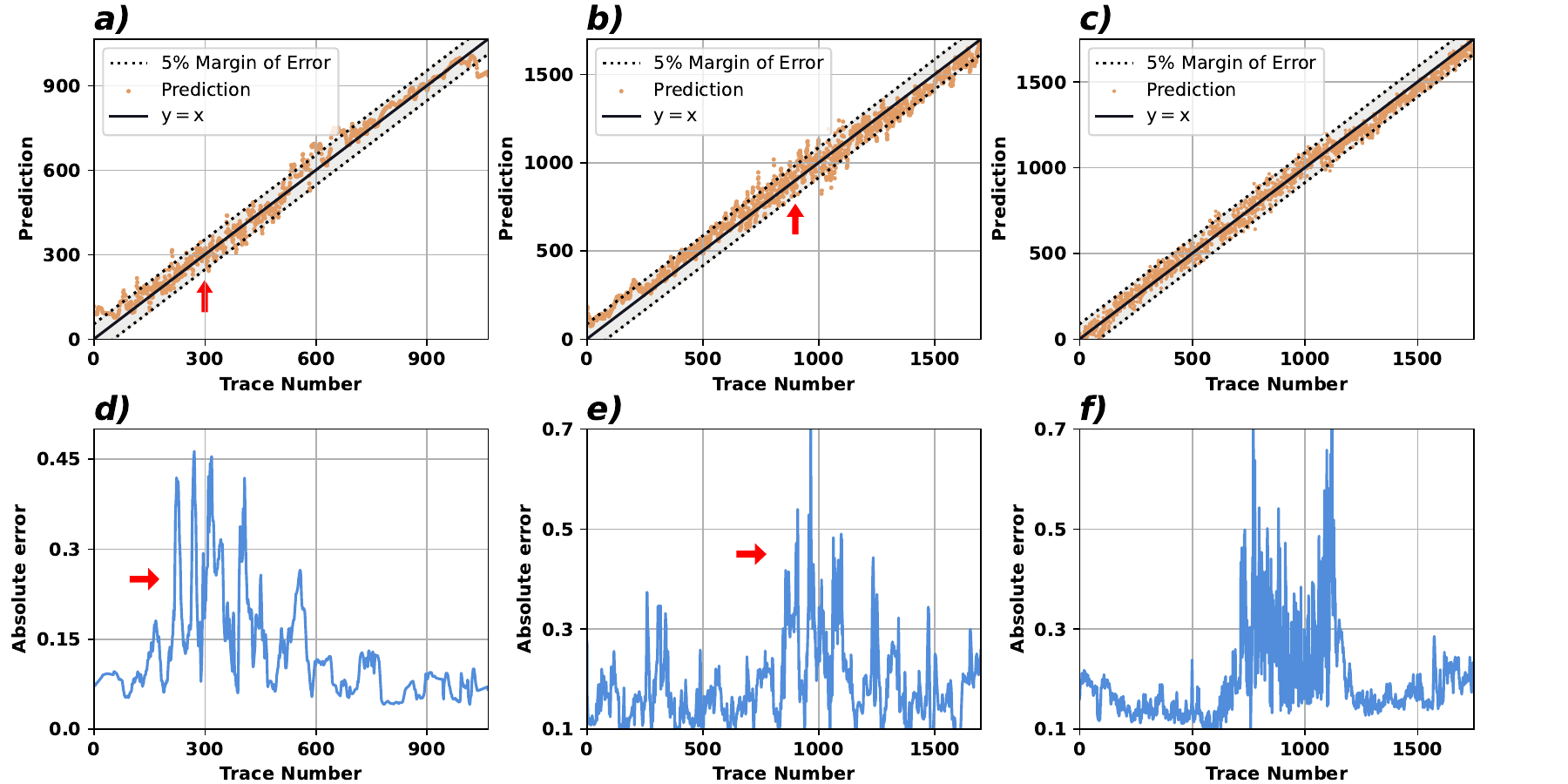}
\caption{a) and d) represent the distribution of each seismic trace of Overthrust after Dimension Reducer, and the absolute error of the inversion result for each seismic trace. b) and e) represent the distribution of each seismic trace of Marmousi 2 after Dimension Reducer, and the absolute error of the inversion result for each seismic trace. c) and f) represent the distribution of each seismic trace of SEAM after Dimension Reducer, and the absolute error of the inversion result for each seismic trace.}
\label{dimension_reduced}
\end{figure}

Deep learning networks constructed with fully connected layers exhibit strong linear generalization ability \citep{49}, making them more suitable for linear feature inputs than for nonlinear ones. Additionally, the Dimension Reducer, through gradient descent, encourages the features encoded by the Encoder to possess contextual properties, effectively providing a positional encoding for each seismic trace. As shown in Figure~\ref{marmousi_whole}, within the region of 10 to 12 km and two-way time of 0 to 1.2 s, the proposed method achieves better horizontal continuity. Furthermore, we statistically analyzed the distribution of each seismic trace after dimension reduction by the Dimension Reducer, as well as the absolute error of the inverted acoustic impedance, as shown in Figure~\ref{dimension_reduced}. It can be observed that the results of the Dimension Reducer do not reach a completely ideal state, with the overall error being less than 5$\%$ and $R^2$ values greater than 0.98 (Overthrust: 0.9808, Marmousi 2: 0.9827, and SEAM: 0.9919). There are two main reasons for these errors: first, in areas with small lateral variations, adjacent seismic traces are highly similar, making it difficult to distinguish the extracted features after dimension reduction; second, in areas with sharp stratigraphic changes, the variations between adjacent seismic traces are too significant to be represented by linear relationships. The former has little impact on the inversion results, as highly similar seismic traces within the same dataset yield highly similar inversion results; the latter, however, significantly affects the linearization of seismic trace features and the extrapolation ability of the inverter, thereby reducing inversion accuracy in these regions. As shown in Figure~\ref{dimension_reduced}d to f, regions with large absolute errors in the inversion results always correspond to regions where the errors of the Dimension Reducer are concentrated (Figure~\ref{dimension_reduced}a to c). From the perspective of various metrics in Table~\ref{results}, linear feature encoding leads to a substantial improvement; however, due to the large number of influencing factors (feature differences between datasets and the inherent complexity of deep learning models), this improvement is difficult to analyze quantitatively. Therefore, further exploration of seismic trace feature encoding remains a highly valuable research direction.

Beyond the Dimension Reducer, the Reconstructor is an essential component—its role is even more critical than that of the Dimension Reducer. The Reconstructor ensures that the extracted features are both meaningful and effective, while the Dimension Reducer enforces linearity in these features. As shown in Table~\ref{heterogeneous}, our experiments demonstrate that omitting the Reconstructor can lead to non-convergence during Encoder and Inverter training, ultimately making the entire AII process infeasible. 

\begin{figure}
\centering
\noindent\includegraphics[width=\textwidth]{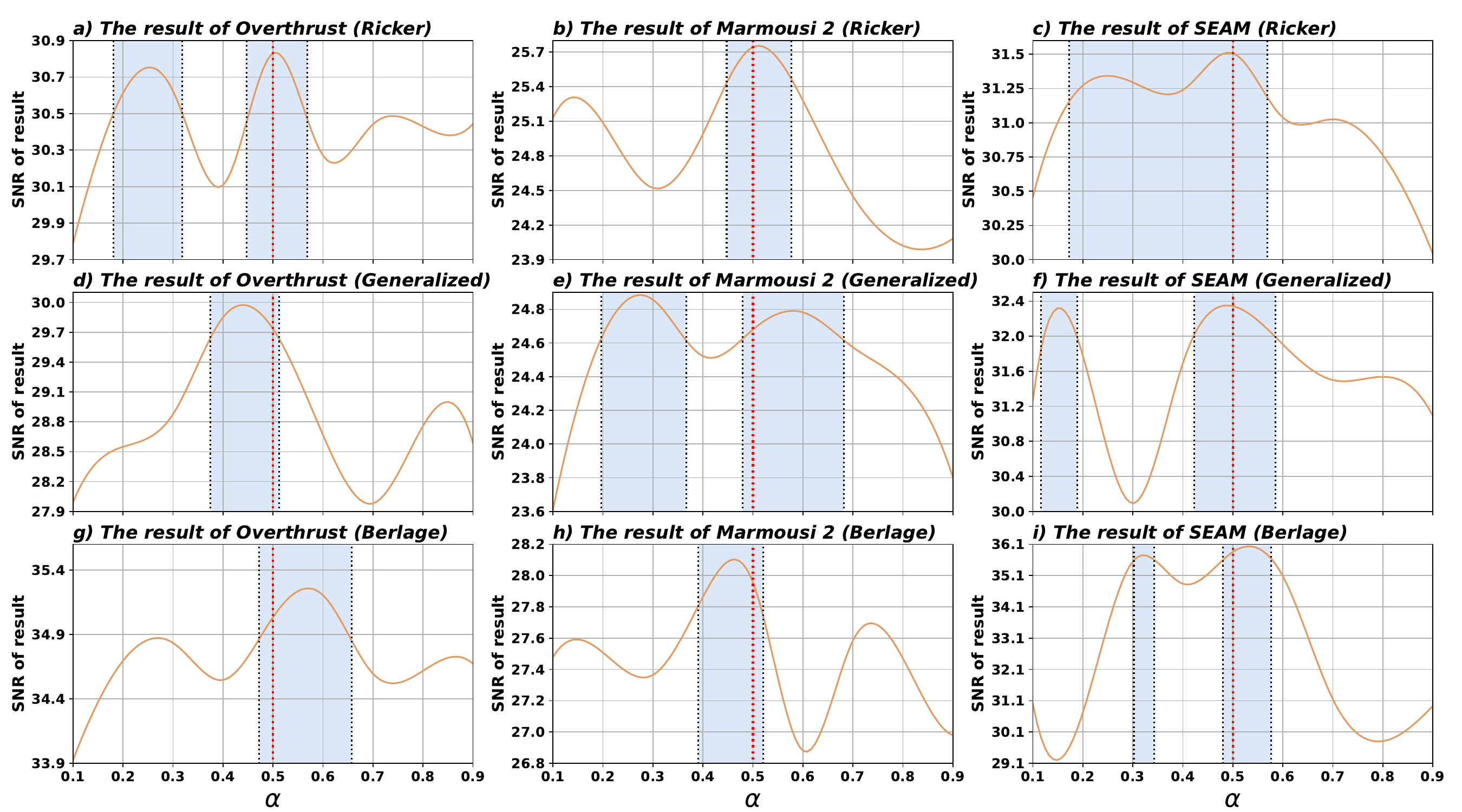}
\caption{Results of the proposed method on each dataset under different $\alpha$ and different wavelets. The blue area represents results whose error from the optimal weight outcome falls within 99$\%$, while the red line indicates the equal weight.}
\label{weight}
\end{figure}

It is also important to note that, during Encoder training, we assigned equal weights to the loss gradients from the Reconstructor and Dimension Reducer. To further examine the impact of this weighting, we introduced a weight factor $\alpha$ to balance the two losses and analyzed its effect on the results, as illustrated in Figure~\ref{weight}. The weight factor $\alpha$ combines the losses as $2 \times[\alpha \times Loss_R + (1 - \alpha)\times Loss_D]$; when $\alpha=0.5$, this corresponds to the equal weighting used in our previous experiments. The results indicate that the optimal value of $\alpha$ varies considerably across different datasets and wavelet types. Although equal weighting is not always optimal, the performance difference compared to the optimal weighting remains within 1$\%$. Due to the large number of influencing factors, it is challenging to determine the optimal weight directly. While heuristic multi-task optimizers could potentially address this issue, they often introduce additional training complexity and reduce overall efficiency.

Structurally, the Encoder is implemented using TCNs, while the Reconstructor and Inverter are built with Bi-GRUs, forming a heterogeneous architecture. Heterogeneous networks were initially popularized in graph neural networks, such as graph convolutional neural networks \citep{65}. In the field of AII, hybrid networks that combine CNNs and GRUs are widely adopted \citep{66}, as they effectively capture the spatiotemporal dynamics and correlations of seismic sequences, resulting in more continuous and stable inversion outcomes \citep{67}. Additionally, the proposed method employs a two-stage training strategy. To accommodate datasets of varying sizes, the designed Encoder does not include any downsampling operations. Consequently, if the Reconstructor adopts the same convolutional structure as the Encoder, the network is prone to shortcut learning \citep{51} (the model may simply learn to copy the input to the output, leading to overfitting), which adversely affects subsequent AII tasks. Therefore, in this work, we designed a heterogeneous structure to mitigate shortcut learning in the Encoder, while also leveraging the complementary strengths of both architectures. To validate this design, we conducted ablation experiments as shown in Table~\ref{heterogeneous}. The results indicate that when the Encoder, Reconstructor, and Inverter all adopt a homogeneous TCN structure, satisfactory performance is difficult to achieve. We speculate that this is due to shortcut learning between the Encoder and Reconstructor during training, which hinders effective training of the Inverter. In contrast, a homogeneous structure based on Bi-GRUs alleviates this issue to some extent, owing to the gating mechanism and increased number of learnable parameters. However, its performance still falls short compared to the heterogeneous architecture. In addition to TCN and Bi-GRU, combining more advanced network architectures such as Transformer \citep{69} also merits further exploration.

\begin{table}[!htbp]
\centering
\tiny
\setlength{\tabcolsep}{1.2mm}
\renewcommand\arraystretch{1.6}
\begin{tabular}{ccccccc}
\toprule[1.5pt]
\textbf{Metric} & \textbf{Dataset}  & \textbf{Heterogeneous} & \textbf{Full TCN} & \textbf{Full Bi-GRU} & \textbf{\makecell{Without \\ Dimension Reducer}} & \textbf{\makecell{Without \\ Reconstructor}} \\
\midrule
\multirow{3}{*}{\textbf{SNR $\uparrow $}} & Overthrust & \textbf{30.8307} & 16.3292 & 28.6502 & 29.5747 & 28.1930 \\
& Marmousi 2 & \textbf{25.7436} & 10.7186 & 24.3773 & 23.9952 & 19.6224 \\
& SEAM & \textbf{31.5077} & 11.1428 & 28.8263 & 30.3412 & 18.0114 \\
\midrule
\multirow{3}{*}{\textbf{$R^2 \uparrow$}} & Overthrust & \textbf{0.9892} & 0.6952 & 0.9821 & 0.9856 & 0.9802 \\
& Marmousi 2 & \textbf{0.9783} & 0.3088 & 0.9702 & 0.9675 & 0.9110 \\
& SEAM & \textbf{0.9944} & 0.3855 & 0.9845 & 0.9926 & 0.8736 \\
\midrule
\multirow{3}{*}{\textbf{SSIM $\uparrow $}} & Overthrust & \textbf{0.9542} & 0.5734 & 0.9392 & 0.9330 & 0.9342 \\
& Marmousi 2 & \textbf{0.9134} & 0.5496 & 0.8907 & 0.8907 & 0.8191 \\
& SEAM & \textbf{0.9277} & 0.4460 & 0.9044 & 0.9131 & 0.8665\\
\midrule
\multirow{3}{*}{\textbf{MAE $\downarrow$}} & Overthrust & \textbf{0.0579} & 0.3936 & 0.0746 & 0.0674 & 0.0822 \\
& Marmousi 2 & \textbf{0.0914} & 0.7565 & 0.1081 & 0.1128 & 0.1723 \\
& SEAM & \textbf{0.0497} & 0.6793 & 0.0594 & 0.0547 & 0.2472 \\
\midrule
\multirow{3}{*}{\textbf{MSE $\downarrow$}} & Overthrust & \textbf{0.0108} & 0.3304 & 0.0179 & 0.0145 & 0.0199 \\
& Marmousi 2 & \textbf{0.0217} & 0.8883 & 0.0299 & 0.0327 & 0.0900 \\
& SEAM & \textbf{0.0055} & 0.7503 & 0.0105 & 0.0072 & 0.1301 \\
\bottomrule[1.5pt]
\end{tabular}
\caption{The quantitative metrics of heterogeneous and homogeneous structures and ablation results on Overthrust, Marmousi 2 and SEAM datasets.}
\label{heterogeneous}
\end{table}

\section{Conclusion}

In this work, we propose a novel Encoder-Inverter framework for seismic acoustic impedance inversion. Two auxiliary models are incorporated to facilitate Encoder training, enabling the mapping of seismic traces into a linear feature space. These linear features, together with a small amount of well-logging data, are then used to fine-tune the Inverter. By transforming the inversion of unlabeled seismic traces into an extrapolation or interpolation problem, the proposed method significantly improves inversion accuracy and robustness. Specifically, the Dimension Reducer enforces the extraction of linear features, while the Reconstructor ensures that these features retain sufficient information for accurate seismic data reconstruction, thereby preserving their effectiveness for subsequent inversion. Comparative experiments demonstrate that the proposed method achieves $R^2$ values of 0.9892, 0.9783, and 0.9944 on the Overthrust, Marmousi 2, and SEAM datasets, respectively, outperforming existing approaches in terms of lateral continuity and inversion resolution. Furthermore, the method yields promising results on field data. To support future research, both the data and code will be made publicly available.

\section{Acknowledgments}
This research is supported by the Chengdu University of Technology Postgraduate Innovative Cultivation Program. The code and examples of this work can be found in \url{https://github.com/lexiaoheng/Encoder-Inverter-Framework-for-Seismic-Acoustic-Impedance-Inversion }. Long-term maintenance versions can be found in \url{https://github.com/lexiaoheng/Mariana}. 

\bibliography{cite}
\bibliographystyle{elsarticle-harv}

\end{document}